\documentclass{pasj01}
\Received{$\langle$reception date$\rangle$}
\Accepted{$\langle$acception date$\rangle$}
\Published{$\langle$publication date$\rangle$}
\begin{document}

\title{The redshift selected sample of long gamma-ray burst host galaxies: 
the overall metallicity distribution at $z < 0.4$\thanks{
Based on observations obtained at the Gemini Observatory 
via the time exchange program between Gemini and the Subaru Telescope 
(processed using the Gemini IRAF package), 
which is operated by the Association of Universities for Research in Astronomy, Inc., 
under a cooperative agreement with the NSF on behalf of the Gemini partnership: 
the National Science Foundation (United States), the National Research Council (Canada), 
CONICYT (Chile), Ministerio de Ciencia, Tecnolog\'{i}a e Innovaci\'{o}n Productiva (Argentina), 
and Minist\'{e}rio da Ci\^{e}ncia, Tecnologia e Inova\c{c}\~{a}o (Brazil).
}\thanks{
Based in part on data collected at Subaru Telescope, 
which is operated by the National Astronomical Observatory of Japan.
}}
\author{
Yuu \textsc{Niino}\altaffilmark{1}, 
Kentaro \textsc{Aoki}\altaffilmark{2}, 
Tetsuya \textsc{Hashimoto}\altaffilmark{1}, 
Takashi \textsc{Hattori}\altaffilmark{2}, 
Shogo \textsc{Ishikawa}\altaffilmark{3}, 
Nobunari \textsc{Kashikawa}\altaffilmark{1}, 
George \textsc{Kosugi}\altaffilmark{1}, 
Masafusa \textsc{Onoue}\altaffilmark{3}, 
Jun \textsc{Toshikawa}\altaffilmark{1}, 
and Kiyoto \textsc{Yabe}\altaffilmark{4} 
}%
\altaffiltext{1}{National Astronomical Observatory of Japan, 
2-21-1 Osawa, Mitaka, Tokyo 181-8588, Japan}
\altaffiltext{2}{Subaru Telescope, National Astronomical Observatory of Japan, 
650 North A‘ohoku Place, Hilo, HI 96720, USA}
\altaffiltext{3}{Department of Astronomy, School of Science, 
SOKENDAI (The Graduate University for Advanced Studies), 
2-21-1 Osawa, Mitaka, Tokyo 181-8588, Japan}
\altaffiltext{4}{Kavli Institute 
for the Physics and Mathematics of the Universe, 
The University of Tokyo, Kashiwanoha, Kashiwa 277-8583, Japan}
\email{yuu.niino@nao.ac.jp}


\maketitle

\begin{abstract}
We discuss the host galaxy metallicity distribution 
of all long gamma-ray bursts (GRBs) whose redshifts are known to be $< 0.4$, 
including newly obtained spectroscopic datasets 
of the host galaxies of GRB 060614, 090417B, and 130427A. 
We compare the metallicity distribution of the 
low-redshift sample to the model predictions, 
and constrain the relation between metallicity and GRB occurrence. 
We take account of spatial variation of metallicities 
among star forming regions within a galaxy. 
We found that the models, in which only low-metallicity stars produce GRBs 
with a sharp cutoff of GRB production efficiency around 12+log(O/H) $\sim$ 8.3, 
can well reproduce the observed distribution, 
while the models with no metallicity dependence are not consistent with the observations.
We also discuss possible sampling biases we may suffer 
by collecting long GRBs whose redshifts are known, 
presenting the photometric observations 
of the host galaxy of GRB 111225A at $z = 0.297$ 
whose redshift has been undetermined until $\sim$ 2.3 years after the burst. 
\end{abstract}

\section{Introduction} 
A long gamma-ray burst (GRB)
is one of the most energetic explosions in the universe, 
which is observed via soft gamma-ray emission 
followed by afterglow in lower energy range. 
It is now broadly agreed that at least some of long GRBs originate 
in core-collapse of massive stars 
together with core-collapse supernovae (CC SNe). 
However, CC SNe do not always accompany long GRBs, 
and the criteria discriminating between long GRBs and general CC SNe 
is one of the most important questions regarding long GRBs. 

Some theoretical studies on the origin of long GRBs 
using stellar evolution models suggest that low-metallicity 
may be a necessary condition for a long GRB to occur 
($Z < \mathrm{a\ few} \times 0.1Z_\odot$, 
e.g. \cite{Yoon:2005a,Hirschi:2005a,Woosley:2006a}). 
Metallicity distribution of long GRB host galaxies provides us with an important clue 
to study the relation between metallicity and a long GRB occurrence. 
Although metallicity of a long GRB host galaxy
is not necessarily identical to that of the progenitor
(e.g., \cite{Niino:2011a,Niino:2011b,Levesque:2011a,Niino:2015a}), 
galaxies with low-metallicity would have higher GRB production efficiency 
[i.e., long GRB rate to star formation rate (SFR) ratio] 
than high-metal galaxies if low-metallicity 
is a necessary condition for a long GRB to occur. 

With a sample of 5 host galaxies (3 long GRBs and 2 X-ray flashes, XRFs), 
\authorcite{Stanek:2006a} (\yearcite{Stanek:2006a}, hereafter S06) showed 
that the metallicity distribution of the host galaxies 
at redshifts $\leq 0.25$ is significantly biased 
towards low-metallicities compared 
with star forming galaxies at similar redshifts, 
suggesting long GRBs really occur in low-metallicity environment. 
However their sample is too small to determine 
the relation between metallicity and long GRB occurrence rate. 
The sample number of spectroscopically studied long GRB host galaxies 
has been dramatically increased during the 10 years after S06 
(e.g., \cite{Savaglio:2009a,Levesque:2010a,Graham:2013a, Kruhler:2015a}). 
However, the relation between metallicity
and long GRB occurrence rate is not understood quantitatively. 

One difficulty is that we can determine redshifts 
and/or identify host galaxies of only a small fraction 
of long GRBs ($\sim$ 30\% in recent observations), 
which is likely biased with respect to the distributions 
of redshifts and host galaxy properties (the $z$-determination/host-identification bias). 
Furthermore, even when the redshift 
and the host galaxy of a long GRB are known, 
the host galaxy is not always studied in detail. 
The targets of the spectroscopic studies have been 
selected non-uniformly by numerous independent researchers 
making the effect of the sampling bias unevaluable 
(the reporting bias, e.g., \cite{Graham:2013a}). 

There have been some previous long GRB host observation 
campaigns invented to overcome the sampling biases 
(so called unbiased surveys, \cite{Hjorth:2012a, Salvaterra:2012a, Perley:2016a}). 
In the unbiased surveys, the samples are selected without using 
informations about redshifts and host galaxies 
to avoid the $z$-determination/host-identification bias, 
and hence the samples span a very wide range of redshifts ($z \sim$ 0–-6). 
In such a wide range of redshift, properties of general galaxies largely evolve 
making it difficult to reach enough statistics to study 
the environmental requirements for a long GRB occurrence 
by comparing the properties of GRB host galaxies to those of general galaxies. 

Some recent studies pick out GRB host galaxies in specific ranges of redshifts 
from the unbiased surveys and compared their properties to those of general galaxies 
at similar redshifts \citep{Schulze:2015a, Vergani:2015a, Perley:2016b, Japelj:2016a}. 
However, the properties of general galaxies, 
to which we compare long GRB host galaxies to investigate 
the environmental requirements for a long GRB occurrence, 
are still not well understood at high-redshifts.  
Metallicity measurement of high-redshift galaxies is an especially challenging issue 
(at $z \gtrsim 1$, e.g., \cite{Kewley:2013a, Maier:2015a}). 

On the other hand, the sample number becomes very small 
at low-redshifts in the unbiased surveys ($< 5$ GRBs at $z < 0.5$) 
because the cosmic long GRB rate density is low 
and the comoving volume element is small, 
although long GRB host galaxies at low-redshifts 
are of special importance to understand the relation between metallicity 
and long GRB occurrence rate as we describe below. 
Hence, although the unbiased surveys are important to understand 
what bias we would suffer by selecting long GRBs 
whose redshifts and host galaxies are known, 
it is difficult to unveil the environmental requirements 
for a long GRB occurrence only with the unbiased surveys. 

Metallicity measurements of the host galaxies of all long GRBs known at low-redshifts 
($z \lesssim 0.5$) can dramatically improve our understanding 
of the relation between metallicity and long GRB occurrence rate. 
There are advantages of studying 
GRB host galaxies at low-redshift in some aspects. 
\begin{enumerate}
\item The success rate of GRB redshift-determinations/host-identifications 
would be higher at lower redshifts (i.e., the $z$-determination/host-identification bias is weaker), 
because faint galaxies can be detected without very deep observations 
and spectroscopies of host galaxies can provide redshift informations 
even when GRB afterglows are not bright enough for spectroscopies. 
\item The reporting bias can be eliminated 
by finding host galaxies of all GRBs known in the redshift range  
and measuring metallicities of all host galaxies, 
which is difficult to do at higher redshifts ($z \gtrsim 0.5$). 
\item A wealth of spectroscopic studies of general galaxies at low-redshifts, 
such as the Sloan Digital Sky Survey (SDSS), 
provides us with control sample of galaxy properties, 
to which we can compare long GRB host properties 
to investigate the relation between long GRB rate and metallicity. 
\end{enumerate}
The advantages and the disadvantages of a redshift selected study 
of GRB host galaxies at low-redshifts can be summarized as in table~\ref{tab:procon}, 
in comparison with the other strategies of GRB host studies. 

In this study, we present the overall metallicity distribution 
of the host galaxies of GRBs known at $z < 0.4$ 
that occurred before the end of March 2014. 
For a few of this low-redshift host galaxies, sufficient spectroscopic data 
to constrain their metallicity were not previously available in the literature, 
or signal-to-noise ratio (S/N) of important 
metallicity indicating emission lines was low. 
We perform spectroscopic metallicity measurements 
of the host galaxies of three low-redshift GRBs, 060614, 090417B, and 130427A, 
to obtain better constraints on the metallicity distribution. 
We compare the metallicity distribution of this low-redshift sample 
to the predictions of empirical models of galaxies, 
and constrain the relation between metallicity and GRB occurrence rate. 

This paper is organized as follows. 
We describe our sample selection, observations, and analysis of the obtained spectra
in section~\ref{sec:sample}, \ref{sec:obs}, and \ref{sec:linefit}, respectively.
In section~\ref{sec:prop}, we discuss the properties 
of the low-redshift long GRB host galaxies including metallicities. 
In section~\ref{sec:moedel}, we describe models of galaxy metallicity distributions, 
to which we compare the low-redshift sample of host galaxies to constrain 
the relation between metallicity and GRB occurrence rate. 
The results of the comparison are presented in section~\ref{sec:result}. 
In section~\ref{sec:discussion}, we discuss uncertainties of our galaxy model, 
possible subpopulations in the low-redshift GRBs, 
and sampling biases we may suffer by collecting long GRBs with known redshifts. 
We summarize our conclusions in section~\ref{sec:conclusion}. 

Throughout this paper, we assume the fiducial cosmology 
with $\Omega_{\Lambda}=0.7$, $\Omega_{m}=0.3$, and $H_0=$ 70 km s$^{-1}$ Mpc$^{-1}$.
The magnitudes are given in the AB system. 

\begin{table}
  \tbl{Strategies of current long GRB host studies}{%
    \begin{tabular}{ccc}
      \hline\noalign{\vskip3pt}
      sampling     & bias                             & sample number\\
                         &                                     & at $z < 0.5$ \\
      \hline\noalign {\vskip1pt} 
      \hline\noalign {\vskip2pt} 
      incomplete  & $z$-determination/host-identification & $\sim 10$ \\
                         & reporting                      & \\
      \hline\noalign{\vskip3pt} 
      unbiased     & none                            & $< 5$ \\
      \hline\noalign{\vskip3pt} 
      redshift       & $z$-determination/host-identification & $\gtrsim 10$ \\
      selected$^\dagger$ & (weaker at lower-redshifts) &  \\
      \hline\noalign{\vskip3pt} 
    \end{tabular}}\label{tab:procon}
  \begin{tabnote}
  $^\dagger$ This work. 
  \end{tabnote}
\end{table}

\section{Sample selection}
\label{sec:sample}

We collect low-redshift GRBs from online databases: 
Swift Gamma-Ray Burst Table\footnote{http://swift.gsfc.nasa.gov/archive/grb\_table/}, 
Gamma-Ray Burst Online Index (GRBOX)\footnote{http://www.astro.caltech.edu/grbox/grbox.php}, 
J. Greiner's GRB table\footnote{http://www.mpe.mpg.de/~jcg/grbgen.html}, 
and GRB Host Studies (GHostS)\footnote{http://www.grbhosts.org/}. 
Although there are various classes of gamma-ray burst like events: 
e.g., long GRBs, short GRBs, ultra-long GRBs, and XRFs, 
we focus on host galaxies of long GRBs in this study. 
The classification of burst events are still controversial 
(e.g., \cite{Zhang:2007a, Bromberg:2013a, Levan:2014a}), 
and important burst properties for the classification 
such as burst duration $T_{90}$, and spectral peak energy $E_{\rm peak}$ 
are uncertain in some cases. 
However, for simplicity, we select long GRBs simply 
by reported $T_{90}$ and $E_{\rm peak}$ as follows. 
The prompt emission properties are taken from
\citet{Butler:2007a}\footnote{http://butler.lab.asu.edu/Swift/}, 
\citet{Troja:2006a}, \citet{Zhang:2009a}, and \citet{Zhang:2012a}, 
in addition to the online databases mentioned above. 

First, we collect bursts with 2 $\leq T_{90} \leq 10^4$ sec as long GRBs, 
considering those with $T_{90} < 2$ sec as short GRBs 
and $T_{90} > 10^4$ sec as ultra-long GRBs. 
We also exclude short GRBs with extended emissions, 
GRB 050709, 050724, 061210, and 071227 from our sample. 
From the bursts which meet the duration criteria, 
we remove bursts with very soft spectra (XRFs) 
which may be a different population from long GRBs. 
XRFs are often defined according to their hardness ratio 
which depends on energy bands of X-ray detectors. 
To take into account burst events discovered by various instruments, 
we define bursts with $E_{\rm peak} < 40$ keV as XRFs in this study. 
For some {\it Swift} detected bursts, $E_{\rm peak}$ estimation is not available. 
In that case, we distinguish GRBs and XRFs using photon spectral index 
in the energy range of {\it Swift}-BAT ($\Gamma_{\rm ph}$).  
We consider bursts with $\Gamma_{\rm ph} \leq 1.8$ 
which is typical of long GRBs with $E_{\rm peak} \geq 40$ keV
\citep{Zhang:2007a, Sakamoto:2009a} as long GRBs, 
and those with $\Gamma_{\rm ph} > 1.8$ as XRFs. 

Within the criteria described above, 15 long GRBs 
occurred before the end of March 2014 at spectroscopically 
measured redshifts $z < 0.4$ (Table~\ref{tab:GRBs}). 
However, we exclude GRB 050219 from further discussion in this study, 
leaving 14 low-redshift long GRBs in our sample, 
because it is unclear whether this burst really has occurred at a low-redshift. 
GRB 050219 occurred close to (but $\sim$ \timeform{2''} offset from) 
an early-galaxy for which the redshift $z = 0.211$ is measured \citep{Rossi:2014a}. 
Given that long GRBs result from death of massive stars, 
the association of GRB 050219 with an early-type galaxy is surprising, 
although it is not strictly prohibited because early-type galaxies 
can harbour small amount of star formation (e.g., \cite{Morganti:2006a}). 
The significant offset of GRB 050219 from its claimed host galaxy 
also suggests that the galaxy might be aligned with the GRB 
by chance without any physical association with the burst. 
It is possible that the remaining sample is also contaminated 
by galaxies that coincide with GRBs by chance. 
We discuss this issue in section~\ref{sec:contamination}. 

Among the 14 GRBs, GRB 060614, 090417B, 111225A, and 130427A 
were without sufficient spectroscopic observations 
to significantly constrain their host metallicity. 
We perform spectroscopic observations 
of the host galaxies of GRB 060614, 090417B, and 130427A in this study. 
There are also some recently published spectroscopic observations 
of the host galaxies of GRB 060614 and 130427A. 
We also discuss those observations in section~\ref{sec:obs}.  

The host galaxy of GRB 111225A missed our spectroscopic observation 
because its redshift had not been measured as of the end of March 2014, 
and no metallicity information is available for this host galaxy. 
The redshift of this burst was reported $\sim$ 2.3 years after the burst \citep{Thone:2014b}. 
We take into account this host galaxy as a part of the uncertainty 
in the metallicity distribution of the GRB host galaxies. 
We discuss the implications for the sampling bias in the redshift selected sample
from the late redshift report of this burst in section~\ref{sec:fainthost}. 
Hereafter, GRB means long GRB unless otherwise stated. 

\begin{table}
  \tbl{Long GRBs at $z < 0.4$}{%
    \begin{tabular}{lrl}
      \hline\noalign{\vskip3pt}
      GRB & redshift & source of redshift$^\dagger$ \\
      \hline\noalign{\vskip3pt} 
      980425$^\ddagger$ & 0.0085 & host galaxy [1] \\
      060505  & 0.089 & host galaxy [2] \\
      080517  & 0.089 & host galaxy [3] \\
      031203$^\ddagger$ & 0.105 & host galaxy [4] \\
      060614  & 0.125 & host galaxy [5, 6] \\
      030329$^\ddagger$ & 0.169 & afterglow, host galaxy [7] \\
      {\em 050219$^{\dagger\dagger}$} & {\em 0.211} & {\em host galaxy}[8] \\
      120422A$^\ddagger$ & 0.283 & afterglow [9], host galaxy [10] \\
      050826 & 0.296 & host galaxy [11] \\
      111225A & 0.297 & afterglow, host galaxy[12] \\
      130427A$^\ddagger$ & 0.340 & afterglow [13], host galaxy [14] \\
      090417B & 0.345 & host galaxy [15]\\
      061021  & 0.346 & afterglow [16], host galaxy [17] \\
      011121$^\ddagger$ & 0.362 & host galaxy [18, 19] \\
      120714B$^\ddagger$ & 0.398 & afterglow, host galaxy [20] \\
      \hline\noalign{\vskip3pt} 
    \end{tabular}}\label{tab:GRBs}
  \begin{tabnote}
  Long GRBs whose redshifts are spectroscopically determined to be $z < 0.4$. 
  Short GRBs, XRFs, and ultra-long GRBs are not included. \\
  $^\dagger$ If `host galaxy', the redshift 
  is determined by emission lines of the host galaxy. 
  If `afterglow', the redshift is determined 
  by absorption lines in the afterglow. 
  The numbers in the square brackets are the references as listed below. \\
  References: 1. \citet{Tinney:1998a}, 2. \citet{Ofek:2006a}, 3. \citet{Stanway:2015a}, 
  4. \citet{Prochaska:2003a}, 5. \citet{Price:2006a}, 6. \citet{Fugazza:2006a}, 
  7. \citet{Greiner:2003a}, 8. \citet{Rossi:2014a}, 9. \citet{Schulze:2012a}, 
  10. \citet{Tanvir:2012a}, 11. \citet{Halpern:2006a}, 12. \citet{Thone:2014b}, 
  13. \citet{Levan:2013a}, 14. \citet{Xu:2013b}, 15. \citet{Berger:2009a}, 
  16. \citet{Fynbo:2009a}, 17. \citet{Hjorth:2012a}, 18. \citet{Infante:2001a}, 
  19. \citet{Garnavich:2003a}, 20. \citet{Fynbo:2012a} \\
  $^\ddagger$ bursts with confirmed SN associations \citep{Cano:2016a} \\
  $^{\dagger\dagger}$ We do not include GRB 050219 in the analyses in this paper, 
  because the association of this bust with its host galaxy 
  for which the redshift is measured is highly uncertain. 
  \end{tabnote}
\end{table}

\section{Observations}
\label{sec:obs}

We perform optical spectroscopy of the host galaxies of GRB 060614, 090417B, and 130427A 
with the Gemini Multi-Object Spectrographs (GMOS, \cite{Hook:2004a}), 
and use flux ratios of the emission lines: 
[O\emissiontype{II}]$\lambda$3727, H$\beta$, [O\emissiontype{III}]$\lambda$4959, 
[O\emissiontype{III}]$\lambda$5007,  H$\alpha$, and [N\emissiontype{II}]$\lambda$6584, as metallicity indicators
(hereafter, [O\emissiontype{II}], [O\emissiontype{III}], and [N\emissiontype{II}] mean 
[O\emissiontype{II}]$\lambda$3727, [O\emissiontype{III}]$\lambda$5007, 
and [N\emissiontype{II}]$\lambda$6584, respectively, unless otherwise stated). 
The apparent spatial extent of the three host galaxies are $\sim$ \timeform{1''},  
and major part of their flux is collected in the slit.   
The spectra obtained with GMOS were calibrated 
using spectroscopic standard stars observed on different nights 
from the observations of the host galaxies, 
and hence the absolute flux scale is not accurate. 

Emission line fluxes of the GRB 130427A host galaxy 
obtained during a target-of-opportunity program (TOO, PI: N. Kawai) 
using Subaru/FOCAS \citep{Kashikawa:2002a} are presented together.  
We also performed imaging observations of the host galaxy of GRB 111225A. 
We describe the observations of each target in the following subsections. 
See table~\ref{tab:obs} for the summary of our observations. 

The data were reduced in a standard manner using 
PyRAF\footnote{PyRAF is a product of the Space Telescope Science Institute, 
which is operated by AURA for NASA}/
IRAF\footnote{IRAF is distributed by the National Optical Astronomy Observatories, 
which are operated by the Association of Universities for Research in Astronomy, Inc., 
under cooperative agreement with the National Science Foundation.}, 
together with the Gemini IRAF package and the FOCASRED package. 

\subsection{Spectroscopy of the GRB 060614 host galaxy}
\label{sec:obs:060614}

GRB 060614 is known as a GRB unassociated with a bright SN at $z = 0.125$
\citep{Gehrels:2006a, Fynbo:2006a, Della-Valle:2006a, Gal-Yam:2006a}. 
Although some spectroscopic observations of the GRB 060614 host galaxy 
were performed soon after the burst \citep{Fynbo:2006a, Della-Valle:2006a, Gal-Yam:2006a}, 
the metallicity measurements were uncertain 
because the fluxes of [N\emissiontype{II}] 
and [O\emissiontype{II}] lines were not well constrained. 
Recently, \citet{Japelj:2016a} also reported 
metallicity measurement of the host galaxy of GRB 060614. 
The metallicity they derived is broadly consistent 
with ours which we discuss in section~\ref{sec:metal}. 

We performed spectroscopy of the GRB 060614 host galaxy with GMOS-South 
using the \timeform{1''.0} slit and two different settings of grisms and order-cut filters. 
One is the B600 grating (no-filter) which covers 3500--6000~\AA, 
and the other is the R400 grating + the OG515 order-cut filter covering 5300--9000~\AA. 
The signals were $2 \times 2$ pixels binned. 
The B600 spectroscopy was performed on 2015 April 25 (UT).  
The integration time was 1000 sec $\times 6$, and 
the slit position angle was set to the parallactic angle to minimize 
the effect of differential atmospheric refraction. 
The R400 spectroscopy was performed on 2015 April 28, and May 14  
with the integration time of 1200 sec $\times 8$ 
and the position angle set to the parallactic angle. 

\subsection{Spectroscopy of the GRB 090417B host galaxy}

GRB 090417B is an optically dark GRB at $z = 0.345$
with a very long duration of $> 2130$ sec \citep{Holland:2010a}. 
The spectroscopy was performed with GMOS-North on 2015 March 25.
We use the \timeform{1''.0} slit, the R831 grating, and the RG610 order-cut filter covering 7300--9200~\AA. 
At this redshift, the H$\alpha$ line coincides with a strong night-sky emission line at 8827~\AA. 
To accurately subtract sky emission lines, we utilize the Nod and Shuffle sky subtraction. 
The signals were 2 pixels binned along the spectral direction. 
The total integration time was 2640 sec 
(60 sec $\times$ 2 positions $\times$ 11 cycles $\times$ 2 sequences). 
The slit position angle was set to the parallactic angle. 

\subsection{Spectroscopy of the GRB 130427A host galaxy}

GRB 130427A is a bright burst at $z = 0.340$ 
associated with a broad-lined Type Ic SN 2013cq \citep{Xu:2013a, Perley:2014a, Levan:2014b}. 
Although detection of the host galaxy emission lines over the afterglow and the SN are reported by \citet{Xu:2013a} 
and \authorcite{Kruhler:2015a}~(\yearcite{Kruhler:2015a}, hereafter K15), 
the detection of the [N\emissiontype{II}] line is marginal in either case. 

The spectroscopic follow up observation of GRB 130427A with Subaru/FOCAS 
was performed on 2013 May 17 (i.e., $\sim$ 20 days after the burst), 
under a weather condition with cirrus clouds (PI: N. Kawai). 
The \timeform{0''.8} slit, the 300B grating, 
and the Y47 filter covering 4700--9000~\AA\ were used. 
The slit position angle was set to \timeform{130D} to put 
both of the center of the host galaxy and the GRB position into the slit. 
The signals were 2 pixels binned along the spatial direction. 
The integration time was 9600 sec. 
Unfortunately the atmospheric diffraction corrector (ADC) 
could not be used at the time of the observation, however 
the spectroscopy was performed with elevation angles $>$ \timeform{60D}, 
and atmospheric refraction would be $\lesssim$ \timeform{0''.25} throughout the wavelength range. 

Although the S/N of the FOCAS spectrum 
is higher than those previously reported, 
the significance of the detection of the [N\emissiontype{II}] line 
was still low $\sim 4\sigma$ ($\sim 2.5\sigma$ in K15). 
Furthermore, the strong night-sky emission line at 8827~\AA\ 
overlaps with the [N\emissiontype{II}] line of the host galaxy at this redshift. 
Hence accurate subtraction of the sky emission is essential 
for the spectroscopy of this host galaxy. 

We performed further spectroscopy 
of the GRB 130427A host galaxy with GMOS-North on 2015 March 27 
to measure the [N\emissiontype{II}] line flux 
more securely with higher significance of the line detection. 
The total integration time was 2640 sec. 
The slit position angle was set to the parallactic angle. 
We used the Nod and Shuffle sky subtraction to subtract the sky emission line accurately. 
The signals were 2 pixels binned along the spectral direction.  

\subsection{Imaging of the GRB 111225A host galaxy}
\label{sec:obs:111225a}

\citet{Thone:2014b} reported the redshift of GRB 111225A, $z=0.297$, in April 2014 
($\sim$ 2.3 years after the burst) by analyzing the archival data with an updated software. 
Thus GRB 111225A was not included in the target list 
of our spectroscopic observations which was compiled in March 2014. 
We perform imaging observations of the host galaxy of GRB 111225A 
in $U-, B-, V-, R-$, and $I-$band filters with Subaru/FOCAS 
on 2015 September 22, under a photometric condition. 
The integration time is 1800 sec for $U-$band, 
and 600 sec for each of $B-, V-, R-$, and $I-$band. 
Signals were $2 \times 2$ pixels binned. 
Photometric calibration is performed with respect to \citet{Landolt:1992a} 
standard stars in the PG0231+051 field observed in the same night. 

\begin{table*}
  \tbl{Observations of the GRB host galaxies}{%
    \begin{tabular}{llllccccc}
      \hline\noalign{\vskip3pt}
      target                     & instrument & grating            & N\&S$^\dagger$ & PA     & slit witdth & Integration                       & resolution & seeing \\
      host galaxy            &                   & + filter            &                           & [deg] & [arcsec]     & [sec]                                 & [R]            & [arcsec] \\
      \hline\noalign{\vskip3pt} 
      GRB 060614           & GMOS-S       & B600                & No                   &  92$^\ddagger$  & 1.0 & 1000 $\times$ 6              & 1300         & 0.9 \\
                                    & GMOS-S       & R400 + OG515 & No                   & 118$^\ddagger$ & 1.0 & 1200 $\times$ 8               & 1000         & 0.7 \\
      GRB 090417B         & GMOS-N      & R831 + RG610 & Yes                  & 122$^\ddagger$  & 1.0 & 1320$^{\dagger\dagger} \times$ 2 & 3000          & 0.6 \\
      GRB 130427A         & GMOS-N      & R831 + RG610 & Yes                  & 160$^\ddagger$ & 1.0 & 1320$^{\dagger\dagger} \times$ 2 & 3000         & 0.4 \\
      \hline\noalign{\vskip3pt} 
      GRB 130427A$^{\ddagger\ddagger}$ & FOCAS & 300B + Y47 & No            & 130    & 0.8           & 1200 $\times$ 8               & 1000       & unstable \\
      \hline\noalign{\vskip3pt} 
      GRB 111225A         & FOCAS        & U (imaging)       & No                  & -      & -                & 360 $\times$ 5                    & -            & 0.8 \\
                                    & FOCAS        & B (imaging)       & No                   & -      & -                & 120 $\times$ 5                    & -            & 0.8 \\
                                    & FOCAS        & V (imaging)       & No                   & -      & -                & 120 $\times$ 5                    & -            & 0.6 \\
                                    & FOCAS        & R (imaging)       & No                   & -      & -                & 120 $\times$ 5                    & -            & 0.7 \\
                                    & FOCAS        & I (imaging)        & No                   & -      & -                & 120 $\times$ 5                    & -            & 0.7 \\
       \hline\noalign{\vskip3pt} 
    \end{tabular}}\label{tab:obs}
  \begin{tabnote}
  The observations of the GRB host galaxies we present in this study. \\
  $^\dagger$ Use of Nod \& Shuffle sky subtraction. \\
  $^\ddagger$ parallactic angle \\
  $^{\dagger\dagger}$ The integration time of 1 Nod \& Shuffle sequence 
  consists of 60 sec $\times$ 2 positions $\times$ 11 cycles. \\
  $^{\ddagger\ddagger}$ TOO, poor weather condition
  \end{tabnote}
\end{table*}

\section{Emission line measurements}
\label{sec:linefit}

\begin{figure*}
 \begin{center}
  \includegraphics[scale=0.5]{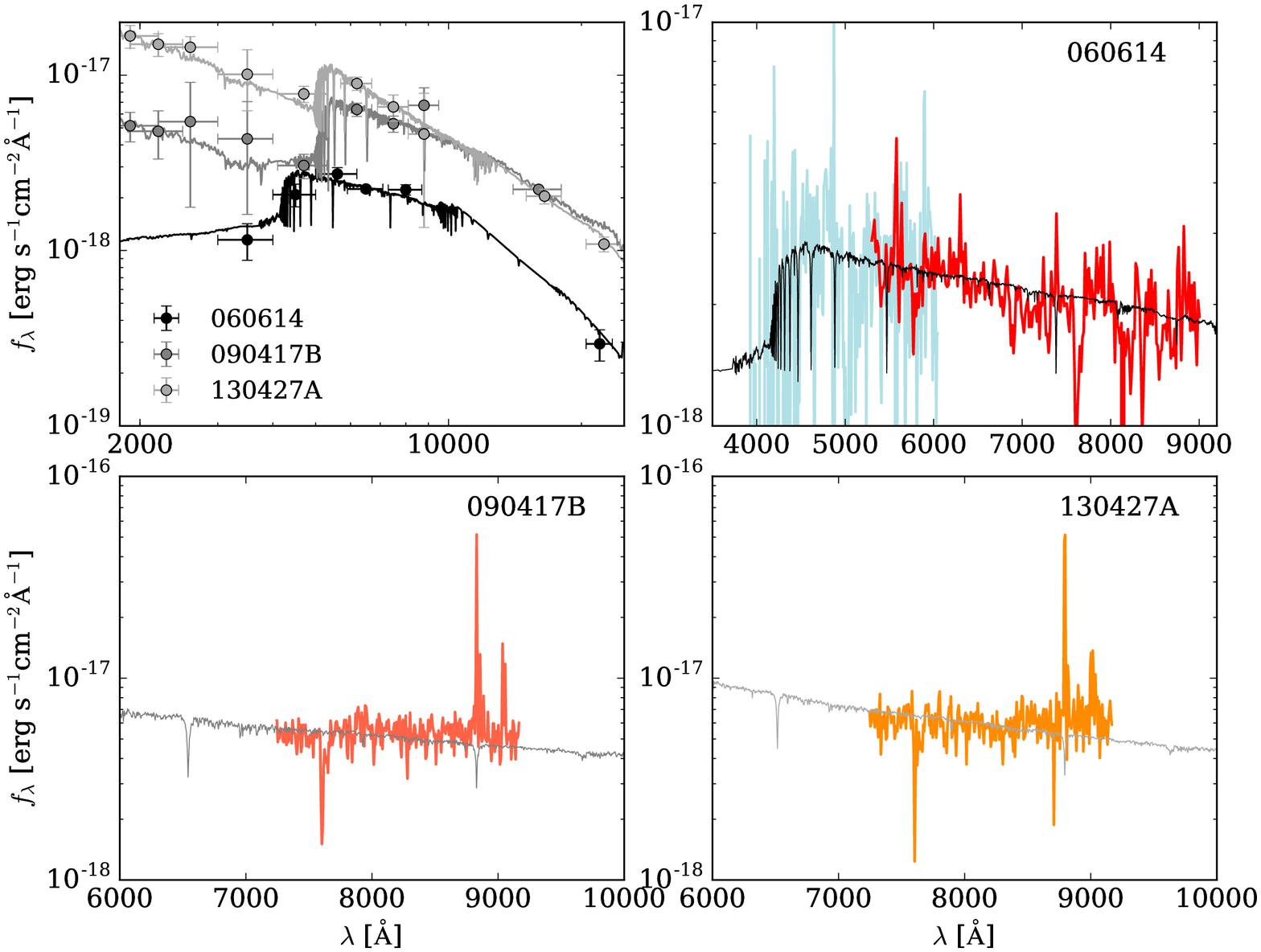}
 \end{center}
 \caption{
   {\it Upper left panel}: the optical/NIR broad-band photometries in the literature 
   and the best fit stellar SED models \citep{Bruzual:2003a}. 
   The photometric data are taken from 
   \citet{Mangano:2007a, Della-Valle:2006a, Hjorth:2012a, Holland:2010a, Perley:2013a, Perley:2014a}, 
   and the SDSS SkyServer (http://skyserver.sdss.org/). 
   {\it Upper right, lower left, and lower right panels}: 
   the best fit stellar SED models (thin lines)
   and the GMOS spectra obtained in this study (thick lines) 
   for the host galaxies of GRB 060614, 090417B, and 130427A, respectively. 
   For the host galaxy of GRB 060614, the spectra obtained with the B600, and R400 gratings 
   are shown with light (blue), and dark (red) colored lines, respectively. 
 }\label{fig:SEDfit}
\end{figure*}

To minimize the effect of stellar absorption features on the line flux measurements, 
we first subtract stellar spectral energy distribution (SED) models 
of the host galaxies from the observed spectra. 
We perform SED fittings to optical/near-infrared (NIR) 
broad-band photometries in the literature (figure~\ref{fig:SEDfit})
using the {\it SEDfit} software package \citep{Sawicki:2012a} 
which utilizes population synthesis models by \citet{Bruzual:2003a}. 
The extinction law by \citet{Calzetti:2000a} 
and the initial mass function (IMF) by \citet{Chabrier:2003a} are assumed. 
We examine six cases of stellar metallicity: $Z_\star = 0.005, 0.02, 0.2, 0.4, 1.0,\ {\rm and}\ 2.5 Z_\odot$, 
and five cases of star formation history: simple stellar population (SSP), constant star formation, 
and exponentially decaying star formation with $\tau = 0.2,\ 1,$ and 5 Gyr. 

As mentioned in section~\ref{sec:obs}, absolute scale of our spectra is not accurate. 
Furthermore, it is possible that a fraction light is lost at the slit. 
In the following discussion, we rescale the observed spectra 
so that the continuum flux agrees with the best fit SED models. 
The broad-band photometries, the best fit SED models, 
and the rescaled spectra are shown in figure~\ref{fig:SEDfit}. 
The parameters of the best fit stellar SED models are listed in table~\ref{tab:pop}. 
We derive metallicities of the GRB host galaxies using relative flux ratios 
between different emission lines as indicators (section~\ref{sec:metal}), 
and hence the derived metallicities are not significantly affected by the error of absolute flux scale. 

\begin{table*}
  \tbl{Results of the host galaxy SED fittings}{%
    \begin{tabular}{lrrrrlr}
      \hline\noalign{\vskip3pt}
      GRB          & ${\rm log_{10}M_\star/M_\odot}$ & Age [Gyr]              & $E(B-V)_{\rm SED}$             & $Z_\star/Z_\odot$              
      & star formation history           & $\chi^2$ \\
      \hline\noalign{\vskip3pt} 
      060614   & $8.2^{+0.5}_{-0.2}$                       & $7.8^{+2.4}_{-0.3}$ & $0.6^{+0.1}_{-0.6}$ & 0.005 & SSP                      & 6.84 \\
      090417B  & $9.4\pm0.1$                            & $8.8\pm0.2$       & $0.2\pm0.1$       & 0.4     & $\tau = 0.2$ Gyr & 4.49 \\
      130427A & $9.0\pm0.1$                             &   $7.8\pm0.2$     & $0.2\pm0.1$       & 0.4     & SSP                        & 1.82 \\
      \hline\noalign{\vskip3pt} 
    \end{tabular}}\label{tab:pop}
  \begin{tabnote}
  The parameters of the best fitting stellar SED models \citep{Bruzual:2003a}
  that reproduce the broad-band photometries in the literature (the upper left panel of figure~\ref{fig:SEDfit}). 
  The extinction law by \citet{Calzetti:2000a} 
  and the IMF by \citet{Chabrier:2003a} are assumed. 
  \end{tabnote}
\end{table*}

There are some archival spectra of the GRB 060614 host galaxy 
as mentioned in section~\ref{sec:obs:060614}, 
and those by \citet{Fynbo:2006a} and \citet{Gal-Yam:2006a} are obtained 
with GMOS-South using the \timeform{1''.0} slit and the R400 grating as well as our R400 spectrum. 
Although the GMOS archival data alone are not deep enough 
to constrain the metallicity of the host galaxy, 
we stack them with our R400 spectra to maximize the S/N. 
The GMOS-South detectors were replaced between 
the archival observations and ours, 
the pixel scale remained $\sim 0.7$ \AA/pix with the R400 grating. 
The integration time of the archival observations 
is 1800 sec $\times$ 4 and 1200 sec $\times$ 4, respectively. 
We note that the detector replacement improved the efficiency 
at the wavelength of H$\alpha$ and [N\emissiontype{II}] at $z = 0.125$ 
by $\sim$ 35\%\footnote{http://www.gemini.edu/sciops/instruments/gmos/imaging/detector-array}. 

\begin{figure*}
 \begin{center}
  \includegraphics[scale=0.5]{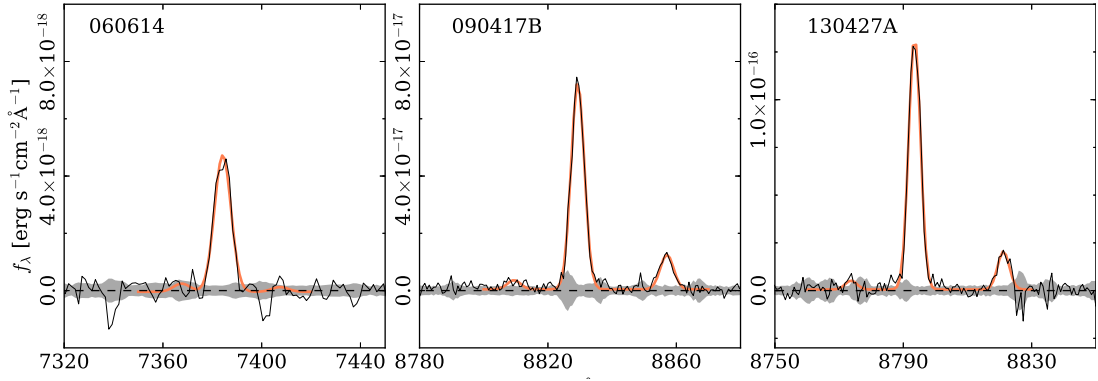}
 \end{center}
 \caption{
   The H$\alpha$ and [N\emissiontype{II}] emission line spectra 
   (the residuals after the model subtraction, thin black lines) 
   of the host galaxies of GRB 060614, 090417B, and 130427A. 
   The thick light-colored (orange) lines are the Gaussian fits 
   to the [N\emissiontype{II}]$\lambda$6548, H$\alpha$, 
   and [N\emissiontype{II}]$\lambda$6584 lines. 
   The shaded regions indicate $\pm 1\sigma$ noise level. 
   For the host galaxies of GRB 060614, the spectrum is stacked 
   with the archival data \citep{Fynbo:2006a, Gal-Yam:2006a}. 
 }\label{fig:fitHa}
\end{figure*}

\begin{figure}
 \begin{center}
  \includegraphics[width=8cm]{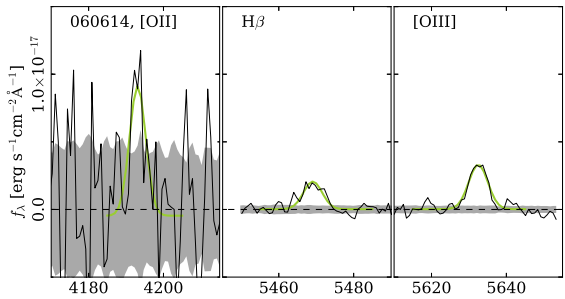}
 \end{center}
 \caption{
   The same as figure~\ref{fig:fitHa}, but for the [O\emissiontype{II}], 
   H$\beta$, [O\emissiontype{III}] lines of the GRB 060614 host galaxy. 
   The spectrum is stacked with the the archival data \citep{Fynbo:2006a, Gal-Yam:2006a} 
   in the wavelength ranges of H$\beta$ and [O\emissiontype{III}]. 
 }\label{fig:fitOxHb}
\end{figure}

We subtract the best fit SED models from the rescaled spectra, 
and measure emission line fluxes by fitting the lines 
with Gaussian plus flat continuum models. 
The residual spectra after the model subtraction and the Gaussian fits 
are shown in figure~\ref{fig:fitHa} and figure~\ref{fig:fitOxHb}. 
Although [O\emissiontype{II}] line is a doublet, 
we fit it with a single Gaussian as well as the other lines, 
because the doublet is not resolved in our spectrum. 
The obtained emission line fluxes are corrected 
for the foreground extinction in the Milky Way (MW)
using the extinction map by \citet{Schlafly:2011a} 
and the extinction law by \citet{Cardelli:1989a}, 
and shown in table~\ref{tab:lines}. 

\begin{figure*}
 \begin{center}
  \includegraphics[scale=0.5]{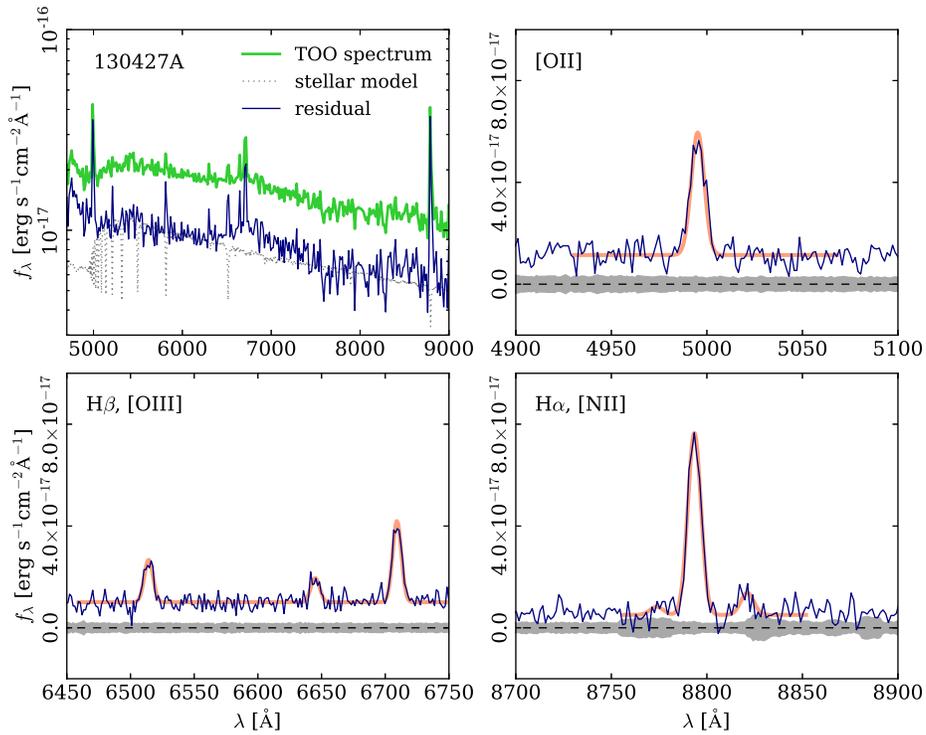}
 \end{center}
 \caption{
   {\it Upper left panel}: the spectrum obtained 
   from the TOO observation of GRB 130427A using Subaru/FOCAS, 
   together with the stellar SED model of the host galaxy 
   (the same as that in the top left panel of figure~\ref{fig:SEDfit}), 
   and the residual spectrum after the model subtraction 
   (i.e., the afterglow, the SN, and the emission lines of the host galaxy). 
   {\it Upper right, lower left, and lower right panels}: 
   the Gaussian plus flat continuum fits to the emission lines in the residual spectrum.  
 }\label{fig:too130427a}
\end{figure*}

The continuum component in the spectrum obtained 
by the FOCAS TOO observation of GRB 130427A 
is contaminated with the afterglow and the SN (figure~\ref{fig:too130427a}). 
Hence we cannot calibrate the flux scale comparing 
the continuum spectrum with the best fit stellar SED model. 
Instead, we firstly measure H$\alpha$ flux without the absorption correction, 
and rescale the FOCAS spectrum so that the absorption uncorrected H$\alpha$ flux 
agrees with that obtained from the GMOS spectrum. 

The emission line fluxes of the other low-redshift GRB host galaxies 
in the literature are also shown in table~\ref{tab:lines}. 
The host galaxies of GRB 980425, 060505, and 120422A, 
are spatially resolved in the spectroscopic observations 
\citep{Christensen:2008a, Thone:2008a, Schulze:2014a}, 
and emission line fluxes are measured at multiple positions within the galaxies. 
In that case, we use the integrated flux over the whole galaxy 
or the flux measured at the peak of the light profile of the host galaxy, 
for consistency with the spectroscopic data 
of the other GRB host galaxies which are not spatially resolved.

\section{Properties of the low-redshift GRB host galaxies}
\label{sec:prop}

\subsection{Metallicity}
\label{sec:metal}

The emission line fluxes of the host galaxies in the low-redshift sample
(see section~\ref{sec:sample}) are listed in table~\ref{tab:lines}. 
With these emission line measurements, 
we derive metallicity of ionized gas, 12+log$_{10}$(O/H), 
which would be close to the metallicity of young stars in the galaxies. 
We utilize both N2 index (=~${\rm log_{10}}$~[N\emissiontype{II}]$\lambda$6584/H$\alpha$) 
and $R_{23}$ index [=~([O\emissiontype{II}]$\lambda$3727+[O\emissiontype{II}]$\lambda$4959+[O\emissiontype{II}]$\lambda$5007)/H$\beta$] 
as indicators of metallicity. 

Various calibrations are proposed for the relation 
between the indiators and metallicity, 
and different calibrations are often inconsistent with each other \citep{Kewley:2008a}. 
In this study, we utilize the method 
of \authorcite{Kobulnicky:2004a}~(\yearcite{Kobulnicky:2004a}, KK04), 
in which both N2 and $R_{23}$ indices 
are calibrated using the same photoionization model, 
to obtain consistent results from the two indicators. 
The solar-metallicity $Z_\odot$ corresponds to 12+log$_{10}$(O/H) = 8.69 \citep{Allende-Prieto:2001a}. 

In the KK04 calibration, the N2 and $R_{23}$ indices are related to metallicity 
taking the effect of ionization parameter ($q$) into account. 
The $q$ parameter can be estimated from $O_{32}$ index 
[=~([O\emissiontype{II}]$\lambda$4959+[O\emissiontype{II}]$\lambda$5007)/[O\emissiontype{II}]$\lambda$3727]. 
The metallicity calibration by KK04 suffers from the model uncertainties of $\sim$ 0.1 dex, 
and hence we consider error of a metallicity measurement to be 0.1 dex 
when the error propagation from the emission line fluxes 
is smaller than this value or flux error informations are not available. 

\begin{figure}
 \begin{center}
  \includegraphics[width=8cm]{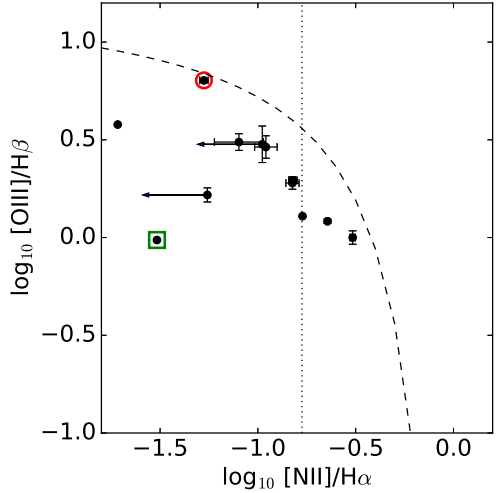}
 \end{center}
 \caption{
   The low-redshift GRB host galaxies on the BPT diagram. 
   The dashed line represents the empirical demarcation 
   between star-forming galaxies and AGNs \citep{Kauffmann:2003a}. 
   The datapoints surrounded by a circle (red) and a square (green) represent 
   the host galaxies of GRB 031203 (with possible AGN, \cite{Levesque:2010c}) and GRB 011121, respectively. 
   N2 and $R_{23}$ indices do not agree well with each other for these host galaxies. 
   The vertical dotted line represents the [N\emissiontype{II}]/H$\alpha$ ratio 
   of the host galaxy of GRB 090417B for which the [O\emissiontype{III}]/H$\beta$ ratio is not known. 
 }\label{fig:BPT}
\end{figure}

The fluxes in table~\ref{tab:lines} are corrected 
for the foreground extinction in the MW, 
but not for the extinction in the host galaxies themselves. 
To derive metallicity, we correct the line fluxes for the host extinctions, 
assuming the intrinsic value of the Balmer decrement H$\alpha$/H$\beta$ = 2.86 
and the extinction law by \citet{Calzetti:2000a}. 
The positions of the host galaxies 
on the Baldwin-Phillips-Terlevich (BPT) diagram \citep{Baldwin:1981a} 
with the extinction correction are shown in figure~\ref{fig:BPT}. 
In the BPT diagram, the GRB host galaxies show emission line ratios 
that are consistent to result purely from star forming activities. 
However, \citet{Levesque:2010c} showed that the spectrum of 
the host galaxy of GRB 031203 is possibly affected by an active galactic nucleus (AGN) 
investigating more numerous emission line fluxes than examined in the BPT diagram. 
Hence the derived properties of the GRB 031203 host galaxy 
might be systematically affected by an AGN. 

We describe the metallicity determination of each host galaxy in the Appendix. 
The derived metallicities and the extinctions 
are listed in table~\ref{tab:metal}~\&~\ref{tab:msfr}. 
When both of the N2 and $R_{23}$ indices are valid for a galaxy, 
we use the $R_{23}$ index, although the two indices agree with each other 
except for the host galaxies of GRB 031203 and 011121. 
The emission line ratio of the GRB 031203 host galaxy 
might be affected by an AGN as mentioned above, 
and the GRB 011121 host galaxy is a remarkable outlier 
from the mass-metallicity ($M_\star$-$Z$) relation of galaxies (see section~\ref{sec:result}). 
The metallicity measurement of these objects should be considered with care. 

\begin{table*}
  \tbl{Emission line fluxes of the GRB host galaxies at $z < 0.4$}{%
    \begin{tabular}{lrrrrrrl}
      \hline\noalign{\vskip3pt}
      GRB         & [O\emissiontype{II}]$\lambda$3727 & H$\beta$ & [O\emissiontype{III}]$\lambda$4959 & 
                        [O\emissiontype{III}]$\lambda$5007 & H$\alpha$ & [N\emissiontype{II}]$\lambda$6584 & ref.\\
      \hline\noalign{\vskip3pt} 
      980425$^\dagger$   & 2.40$\times 10^{4}$ & 4.45$\times 10^{3}$ & -                                & 1.29$\times 10^{4}$ & 1.79$\times 10^{4}$ & 1.97$\times 10^{3}$ & \citet{Christensen:2008a} \\
      060505                   & 16.8$\pm$0.4          & 4.27$\pm$0.03        & 2.19$\pm$0.13        & 5.17$\pm$0.13        & 21.9$\pm$0.09        & 4.97$\pm$0.09        & \citet{Thone:2008a} \\
      080517                   & 1100$\pm$110        & 380$\pm$22            & 130$\pm$6.7           & 380$\pm$22            & 1900$\pm$44          & 580$\pm$13            & \citet{Stanway:2015a} \\
      031203                   & 2270$\pm$230        & 2140$\pm$40          & 4520 $\pm$50         & 13630$\pm$70        & 6040$\pm$30          & 320$\pm$10            & \citet{Prochaska:2004a}; \\
                                                                                                                                                                                                                                                &&&&&&& \citet{Sollerman:2005a} \\
      060614                   & 5.60$\pm$1.16        & 1.35$\pm$0.10        & -                               & 2.23$\pm$0.10        & 4.17$\pm$0.09        & $<$ 0.28                  & {\bf This work}\\
      030329$^\ddagger$  & 64.8                          & 47.3                          & 57.8                         & 179                           & 151                           & 2.9                            & \citet{Levesque:2010c} \\ % no error
      120422A                  & 58.0$\pm$6.7           & 12.8$\pm$0.4          & 8.3$\pm$0.3           & 25.1$\pm$0.5          & 53.6$\pm$0.5          & 8.1$\pm$0.4            & \citet{Schulze:2014a} \\
      050826$^\ddagger$  & 70.9                          & 24.5                          & 10.8                         & 31.5                          & 75.7                          & 12.8                          & \citet{Levesque:2010a} \\ % no error
      111225A                 & -                               & -                               & -                              & -                               & -                                & -                              & - \\
      130427A                 & 43.7$\pm$1.8          & 15.0$\pm$1.0          & 8.5$\pm$1.0            & 28.5$\pm$1.0          & 62.3$\pm$0.9           & 9.36$\pm$0.74        & {\bf This work}$^{\dagger\dagger}$ \\
                                     & (20.2$\pm$2.3           & 4.8$\pm$0.9            & $<$ 2.2                    & 6.5$\pm$1.0            & 14.6$\pm$1.3          & 1.5$\pm$0.6             & K15)\\
      090417B                  & -                               & -                               & -                              & -                               & 38.1$\pm$0.6           & 6.40$\pm$0.51        & {\bf This work}\\
      061021                    & 1.6$\pm$0.1            & 0.5$\pm$0.1            & 0.6$\pm$0.1            & 1.5$\pm$0.2            & 1.9$\pm$0.1            & $<$ 0.26                   & K15\\
      011121$^\ddagger$  & 57.97                       & 17.13                         & 16.79                        & 16.64                       & 65.61                        & 2.0                            & \citet{Garnavich:2003a}; \\
                                                                                                                                                                                                                                                &&&&&&& \citet{Graham:2013a}\\ % no error
      120714B                  & 9.8$\pm$0.5            & 2.5$\pm$0.2             & 2.9$\pm$0.3            & 7.7$\pm$0.5           & 7.5$\pm$0.3             & 0.6$\pm$0.2            & K15\\
      \hline\noalign{\vskip3pt} 
    \end{tabular}}\label{tab:lines}
  \begin{tabnote}
  Fluxes are in units of $10^{-17}$ erg s$^{-1}$cm$^{-2}$, 
  and corrected for the foreground extinction in the MW. 
  When the estimated flux is consistent with zero within the 2$\sigma$ error, 
  we show the 95\% upper-limits instead. \\
  $^\dagger$ The flux errors are $\sim$ 10\%.\\ $^\ddagger$ Flux error informations are not available. \\
  $^{\dagger\dagger}$ The H$\alpha$ and [N\emissiontype{II}] fluxes are measured from the GMOS spectrum, 
  and the other line fluxes are measured from the FOCAS spectrum. 
  \end{tabnote}
\end{table*}

\begin{table*}
  \tbl{Metallicity mesuremetns of the host galaxies}{%
    \begin{tabular}{lrrrr}
      \hline\noalign{\vskip3pt}
      GRB        & \multicolumn{3}{c}{12+log$_{10}$(O/H)$^{ab}$} & log$_{10}\ q^a$ \\
      \cline{2-4}
                    & N2                & $R_{23}$               & best$^{c}$   & \\
      \hline\noalign{\vskip3pt} 
      980425   & $8.57$        & -                          & $8.57$        & 7.3 \\
      060505   & $8.72$        & -                          & $8.72$        & 7.2 \\
      080517   & $8.83$        & $8.68/8.38$        & $8.68$        & 7.2 \\
      031203$^d$ & $8.75$   & $8.48/8.32$        & $8.48$        & 8.3 \\
      060614   & $<8.4$        & $8.66/8.35$        & $8.35$        & 7.3 \\
      030329   & $8.07$        & $8.75/8.14$        & $8.14$        & 7.8 \\
      120422A & $8.65$        & -                          & $8.65$        & 7.3 \\
      050826   & $8.76$        & $8.86/8.14$        & $8.86$        & 7.5 \\
      111225A & -                 & -                          & -                 & -   \\
      130427A & $8.71$        & $8.67/8.33$        & $8.67$        & 7.4 \\
                     & ($8.51\pm0.11$ & 8.70$^{+0.11}_{-0.14}$/8.34$^{+0.11}_{-0.14}$ & - & 7.3--7.2)$^e$ \\
      090417B & 8.6--8.9     & -                          &  8.6--8.9    & 7.2--7.8$^f$ \\
      061021   & $<8.7$       & 8.51$^{+0.15}_{-0.19}$/8.43$^{+0.14}_{-0.15}$   & 8.28--8.66 & 7.5 \\
      011121   & $8.00$        & $8.69/8.33$        & $8.33$        & 7.2 \\
      120714B & $8.54$        & $8.50/8.43$        & $8.50$        & 7.5 \\
      \hline\noalign{\vskip3pt} 
    \end{tabular}}\label{tab:metal}
  \begin{tabnote}
  $^a$ KK04 calibration method is utilized. \\
  $^b$ The estimation errors are $\sim\pm$ 0.1 unless otherwise stated. \\
  $^c$ The best estimates as described in the Appendix. \\ 
  $^d$ Possibly contaminated by an AGN. \\ 
  $^e$ Derived from the emission line fluxes by K15. \\ 
  $^f$ assumed value (see the Appendix) \\
  \end{tabnote}
\end{table*}

\subsection{Stellar mass and SFR}
\label{sec:host_msfr}

Stellar mass ($M_\star$) and SFR of the low-redshift GRB host galaxies are listed in table~\ref{tab:msfr}. 
The stellar masses of the host galaxies of GRB 060614, 090417B, and 130427A 
are derived by the SED fitting assuming the Chabrier IMF as described in section~\ref{sec:linefit}, 
and are broadly consistent with previously reported results \citep{Savaglio:2009a, Xu:2013a, Perley:2013a}. 
The GRB 111225A host galaxy is detected only in 4 bands 
($B, V, R$, and $I$, see table~\ref{tab:111225a}), 
and it is difficult to derive the properties of this host galaxy from its SED. 
However, we can naively estimate its stellar mass as  
$M_\star \sim 4\times10^8M_\odot$ assuming typical mass-to-luminosity ratio 
of galaxies at similar redshifts \citep{Kauffmann:2003b}. 
We collect $M_\star$ measurements of the other host galaxies from the literature. 
When $M_\star$ in the literature is derived assuming an IMF other than the Chabrier IMF, 
we rescale the mass assuming: $M_{\rm \star,Chabrier} = 0.56M_{\rm \star,Salpeter} = 0.83M_{\rm \star,Kroupa}$
\citep{Chabrier:2003a, Brinchmann:2004a}, for consistency. 

We derive SFR of the host galaxies from their extinction corrected H$\alpha$ fluxes 
assuming the global Kennicutt-Schmidt law \citep{Kennicutt:1998a} 
and the same scaling factor between the different IMFs as mentioned above. 
We note that major part of the errors of SFR 
is propageted from the error of $E(B-V)$ in most of the cases. 
The host galaxy of GRB 060505 is spatially extended over a few $\times$ 10 arcsec, 
and only a part of the emission line fluxes is collected in the slit, 
and hence we adopt a SFR measured from ultraviolet photometries 
without dust correction instead \citep{Castro-Ceron:2010a}. 
Although the dust-uncorrected SFR is much larger 
than that suggested from the H$\alpha$ flux with the large slit-loss, 
the actual SFR with dust correction might be further high.
The emission lines of the GRB 120714B host galaxy 
were detected over the optical afterglow (\cite{Fynbo:2012a}; K15), 
and the fraction of the host galaxy light in the slit is not known. 
Hence we consider the obtained SFR 
for the host galaxies of GRB 060505 and 120714B  as lower-limits. 

\begin{table}
  \tbl{Photometries of the host galaxy of GRB 111225A}{%
    \begin{tabular}{rrrrr}
      \hline\noalign{\vskip3pt}
      $U$                     & $B$                  & $V$                  & $R$                 & $I$ \\
      \hline\noalign{\vskip3pt} 
      $> 24.5^\dagger$ & $25.3\pm0.2$ & $24.5\pm0.1$ & $24.7\pm0.1$ & $24.1\pm0.1$ \\
      \hline\noalign{\vskip3pt} 
    \end{tabular}}\label{tab:111225a}
  \begin{tabnote}
  Magnitudes are in the AB system, and not corrected 
  for the foreground extinction in the MW [$E(B-V)=0.23$, \cite{Schlafly:2011a}]. \\
  $^\dagger$ 2$\sigma$-limit. 
  \end{tabnote}
\end{table}

\begin{table}
  \tbl{Dust extinction, stellar mass and SFR of the host galaxies}{%
    \begin{tabular}{lrrr}
      \hline\noalign{\vskip3pt}
      GRB        & $E(B-V)_{\rm BD}^a$ & ${\rm log_{10}}M_\star/M_\odot^b$ & SFR [$M_\odot$yr$^{-1}$]$^{bc}$ \\
      \hline\noalign{\vskip3pt} 
      980425   & 0.29$\pm$0.11 & 9.22$\pm$0.52 [1]     & 0.3$\pm$0.1 \\
      060505   & 0.50$\pm$0.01 & 9.64$\pm$0.02 [2]     & $>$ 1.1 [3] \\
      080517   & 0.48$\pm$0.05 & 9.50$^{+0.12}_{-0.16}$ [4] & 7.2$^{+1.2}_{-1.0}$ \\
      031203$^d$ & $<$ 0.01      & 8.26$\pm$0.45 [1]     & 7.5$\pm$0.2 \\
      060614   & 0.07$\pm$0.06 & $8.2^{+0.5}_{-0.2}$ [5]   & $(9.4^{+1.9}_{-1.6})\times10^{-3}$ \\
      030329   & 0.1                     & 7.91$^{+0.12}_{-0.44}$ [1] & 0.71 \\
      120422A & 0.33$\pm$0.03 & 8.95$\pm$0.04 [6]    & 1.65$\pm$0.15 \\
      050826   & 0.07                   & 10.10$^{+0.22}_{-0.26}$ [1] & 1.17 \\
      111225A & -                        & $\sim$ 8.6 [5]           & - \\
      130427A & 0.32$\pm$0.06 & $9.0\pm0.1$ [5]        & 2.82$^{+0.53}_{-0.45}$ \\
      090417B & -                         & $9.4\pm0.1$ [5]        & 1.25$^{+0.45e}_{-0.33}$ \\
      061021   & 0.24$\pm$0.16 & $8.5\pm0.5$ [7]        & 0.07$^{+0.04}_{-0.03}$ \\
      011121   & 0.25                   & 9.67$\pm$0.17 [8]    & 2.8 \\
      120714B & 0.04$\pm$0.07 & -                                 & $>$ 0.2 \\
      \hline\noalign{\vskip3pt} 
    \end{tabular}}\label{tab:msfr}
  \begin{tabnote}
  $^a$ estimated from the Balmer decrement \\ 
  $^b$ The Chabrier IMF is assumed. The numbers in the square brackets are references as listed below. \\
  $^c$ SFR is derived from the H$\alpha$ flux in table~\ref{tab:lines} 
  and $E(B-V)_{\rm BD}$ in the second column unless otherwise stated. \\
  $^d$ Possibly contaminated by an AGN. \\ 
  $^e$ The extinction correction is based on the result of the SED fitting: $E(B-V)_{\rm SED} = 0.2\pm0.1$.  \\ 
  References: 1. \citet{Levesque:2010a}, 2. \citet{Thone:2008a}, 3. \citet{Castro-Ceron:2010a}, 4. \citet{Stanway:2015a}, 
  5.  This work, 6. \citet{Schulze:2014a}, 7. \citet{Vergani:2015a}, 8. \citet{Kupcu-Yoldas:2007a}
  \end{tabnote}
\end{table}

\section{Models of the metallicity distribution}
\label{sec:moedel}

\subsection{Global properties of star-forming galaxies}

To investigate the relation between metallicity and GRB occurrence, 
we compare the metallicity distribution 
of the low-redshift sample presented 
in section~\ref{sec:prop} with the metallicity distribution 
of star forming galaxies at similar redshifts. 
Following \citet{Stanek:2006a} and \citet{Niino:2011b}, 
we compute the metallicity distribution of star forming galaxies 
using the empirical formulation of stellar mass function [$\phi(M_\star)$], 
$M_\star$-SFR relation (so called ``galaxy main-sequence''), 
and $M_\star$-$Z$ relation of galaxies. 
In this study, we assume $\phi(M_\star)$ by 
\authorcite{Baldry:2012a}~(\yearcite{Baldry:2012a}, blue population), 
the $M_\star$-SFR relation by 
\authorcite{Salim:2007a}~(\yearcite{Salim:2007a}, 
see their equation~11, scatter $\sigma_{\rm MSFR} = 0.5$ dex),  
and the $M_\star$-$Z$ relation computed with the KK04 calibration by \citet{Kewley:2008a}. 

We only consider the mass range log$_{10} M_\star \geq$ 8.0, 
below which the properties of galaxies are not well constrained. 
This mass limit is also comparable 
to the lowest-mass GRB host galaxies in the low-redshift sample. 
$M_\star$ and SFR in the formulations are rescaled to be consistent with the Chabrier IMF. 
The scatter of the $M_\star$-$Z$ relation is largely dependent on $M_\star$ 
in the sense that the scatter is larger at smaller $M_\star$. 
Here we assume a simple formulation: 
\begin{eqnarray}
\sigma_{\rm MZ} = \left\{
  \begin{array}{lr}
    -0.05 {\rm log_{10}}M_\star + 0.6 & ({\rm log_{10}}M_\star < 10.5) \\
    0.075 & ({\rm log_{10}}M_\star \geq 10.5)
  \end{array}\right., \label{eq:sigmaMZ}
\end{eqnarray}
which is broadly consistent with the scatter 
of the $M_\star$-$Z$ relation discussed in \citet{Tremonti:2004a}. 

It should be noted that \citet{Kewley:2008a}  derived the $M_\star$-$Z$ relation 
using galaxy spectra obtained by the SDSS whose fiber spectrograph 
covers only the central \timeform{3''} of each galaxy. 
Hence it is possible that the systematic loss of light 
from the outskirts of galaxies affects the derived metallicities, 
because galaxies often have lower metallicity in their outskirts 
than at their centers (so called metallicity gradient, e.g., \cite{Shields:1978a}). 
Some of the low-redshift GRB host galaxies might also suffer from similar problems. 
However, \citet{Niino:2012a} examined correlation between 
measured metallicities and fraction of light covered by the spectrograph 
using a sample of galaxies with similar $M_\star$ and SFR in the SDSS, 
and found that the dependence of derived metallicities 
on the fiber coviring fraction is small ($< 0.03$ dex, see figure~5 of \cite{Niino:2012a}). 
Thus we consider that the systematic loss of light from the outskirts of galaxies 
does not significantly affect our metallicity measurements

Although it is suggested that there is a correlation between SFR and metallicity of galaxies 
with similar stellar masses \citep{Ellison:2008a, Mannucci:2010a, Lara-Lopez:2010a}, 
the observed correlation between SFR and $Z$ correlation is 
not quantitatively understood and possibly affected by sample selections 
and metallicity calibration methods 
(e.g. \cite{Yates:2012a, Niino:2012a, Andrews:2013a}). 
Hence we do not consider the correlation in our baseline model, 
and discuss the effects of the SFR-$Z$ correlation on our results in section~\ref{sec:MSFRZ}. 

In this study, we consider normalized probability distribution function (PDF) 
of GRB host galaxy metallicities when we compare the model predictions to the observations. 
We note that the absolute scale of the stellar mass function ($\phi^*$), 
and also the absolute scale and the scatter of the $M_\star$-SFR relation 
do not affect the predicted metallicity PDF. 
\citet{Niino:2011b} have shown that the choice of the formulations 
of the stellar mass function and the $M_\star$-SFR relation does not 
significantly affect the predicted metallicity PDF, 
as far as the stellar mass function and the $M_\star$-SFR relation 
that represent field star forming galaxies are selected. 

\subsection{Internal metallicity variation within a galaxy}
\label{sec:Zint}

Observations of some nearby galaxies (MW, the Magellanic Clouds, and M31) 
show that inter-stellar medium (ISM) in a galaxy is not chemically homogeneous. 
Hence the metallicity of a GRB progenitor star 
might be different from the metallicity of the host galaxy 
which we can measure by follow up spectroscopies. 

To examine metallicity variation of star forming regions within a galaxy, 
we consider observed metallicity distributions 
of H\emissiontype{II} regions in nearby galaxies. 
Although recent integral field unit (IFU) spectroscopies provides us 
with spatially resolved map of metallicity in some star forming galaxies, 
the spatial resolution of such observations 
are still limited to $\gtrsim$ 100 pc in most of the cases (e.g., \cite{Sanchez:2012a}). 
Hence the spectroscopy of individual H\emissiontype{II} regions 
(except for giant ones) can be performed only in a few nearby galaxies. 

\begin{figure}
 \begin{center}
  \includegraphics[width=8cm]{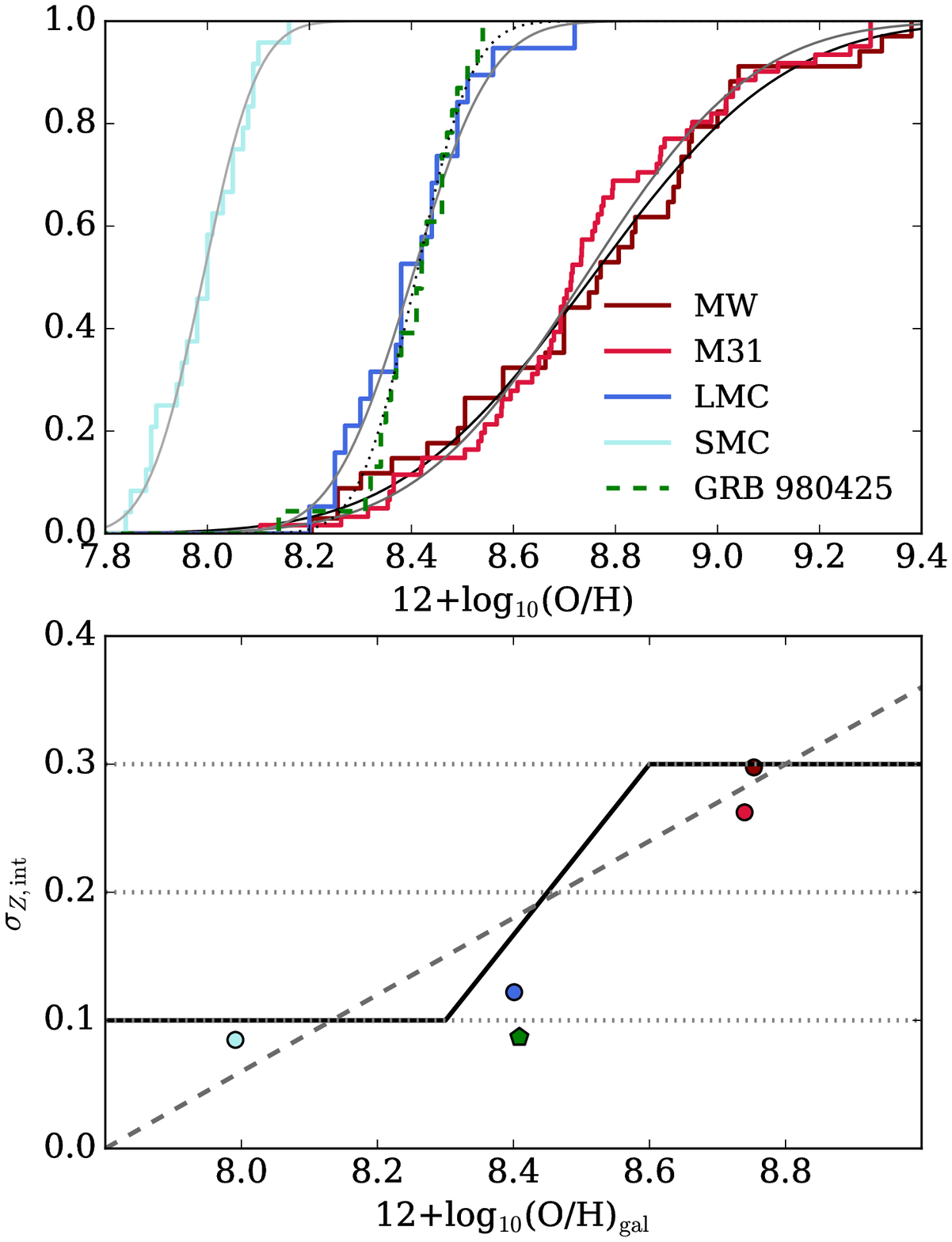}
 \end{center}
 \caption{
   {\it Upper panel}: the metallicity distributions of H\emissiontype{II} regions 
   in the MW (soler-neighbourhood, \cite{Afflerbach:1997a}), 
   the Magellanic clouds \citep{Pagel:1978a}, M31 
   (\cite{Sanders:2012c}, bright objects with $L_{\rm H\alpha} > 10^{36.5}$ [erg s$^{-1}$]), 
   and the host galaxy of GRB 980425 which resides at $z = 0.0085$. 
   The best fit log-normal distributions are plotted together. 
   {\it Lower panel}: the median [12+log$_{10}$(O/H)$_{\rm gal}$] 
   and the scatter ($\sigma_{Z,{\rm int}}$) of the H\emissiontype{II} region 
   metallicity distributions in the nearby galaxies. 
   The pentagonal symbol represents the GRB 980425 host galaxy. 
   The correlations between 12+log$_{10}$(O/H)$_{\rm gal}$ and $\sigma_{Z,{\rm int}}$ 
   which we examine in this study: a broken linear model (solid line), 
   a single linear model (dashed line), and constant models (dotted lines) are plotted together. 
 }\label{fig:Zint}
\end{figure}

In the upper panel of figure~\ref{fig:Zint}, 
we show the metallicity distributions of H\emissiontype{II} regions
in the MW (solar-neighborhood, \cite{Afflerbach:1997a}), 
the large/small Magellanic clouds (LMC and SMC, \cite{Pagel:1978a}), 
and the Andromeda galaxy (M31, \cite{Sanders:2012c}). 
It is possible that the metallicity estimates 
of the faint H\emissiontype{II} regions in the M31 sample 
by \citet{Sanders:2012c} have systematic errors \citep{Niino:2015a},  
and hence we consider only H\emissiontype{II} regions 
with H$\alpha$ luminosity $> 10^{36.5}$ [erg s$^{-1}$] in the M31 sample. 
The H$\alpha$ luminosities of the H\emissiontype{II} regions 
are taken from \citet{Azimlu:2011a}. 
See \cite{Niino:2015a} for the detail of the catalog matching 
between the samples of \citet{Azimlu:2011a} and \citet{Sanders:2012c}. 

The metallicities of the H\emissiontype{II} regions can be represented by log-normal distributions 
but with different median value and dispersion in different galaxies. 
The median and the scatter of the H\emissiontype{II} region metallicity distributions 
in the nearby galaxies [12+log$_{10}$(O/H)$_{\rm gal}$ and $\sigma_{Z,{\rm int}}$]
are shown in the lower panel of figure~\ref{fig:Zint}. 
Although the sample number is small, 
it is seen that the galaxies with the higher median metallicity 
have larger scatter of the distribution. 
As a baseline model, we assume a broken linear relation 
between 12+log$_{10}$(O/H)$_{\rm gal}$ and $\sigma_{Z,{\rm int}}$ 
as shown in the lower panel of figure~\ref{fig:Zint}: 
\begin{eqnarray}
\sigma_{Z,{\rm int}} = \left\{
  \begin{array}{lr}
    0.1 & ({\rm log_{10}(\frac{O}{H})_{gal}} < -3.7) \\
    \frac{2}{3}\times[{\rm log_{10}(\frac{O}{H})_{gal}} & \\
    \ \ \ \ \ \ \ \ + 3.85] & (-3.7 \leq {\rm log_{10}(\frac{O}{H})_{gal}} < -3.4) \\
    0.3 & -3.4 (\leq {\rm log_{10}(\frac{O}{H})_{gal}}) \\
  \end{array}\right., 
\end{eqnarray} 
considering 12+log$_{10}$(O/H)$_{\rm gal}$ equals to the representative metallicity 
of a galaxy which we obtain by spectroscopically observing the galaxy without spatially resolving it.  
We also discuss the results with different $\sigma_{Z,{\rm int}}$ 
formulations in section~\ref{sec:sigma_Zint}. 

In figure~\ref{fig:Zint}, we plot the metallicity distribution 
of H\emissiontype{II} regions in the GRB 980425 host galaxy 
together with those in the local galaxies.  
The GRB 980425 host galaxy is the nearest GRB host galaxy known 
and the only GRB host galaxy for which metallicity 
is measured with a spatial resolution $< 1$ kpc, 
although the resolution $\sim 400$ pc is much larger 
than that achieved for the local galaxies (e.g., $\sim 5$ pc for M31, \cite{Sanders:2012c}). 
The internal metallicity variation observed within the GRB 980425 host galaxy 
is naively consistent with those seen in the Magellanic clouds. 

Metallicity distribution of young stars which are formed in a galaxy 
might be significantly different from the metallicity distribution 
of H\emissiontype{II} regions in the galaxy, 
if metallicity and SFR of H\emissiontype{II} regions are correlated with each other. 
\citet{Niino:2015a} examined the correlation 
between H$\alpha$ luminosity and metallicity of H\emissiontype{II} regions in M31 
using the narrow band photometric data by \citet{Azimlu:2011a} 
and the spectroscopic data by \citet{Sanders:2012c}. 
Although it is suggested that metallicity measurements 
of faint H\emissiontype{II} regions with H$\alpha$ luminosity $< 10^{36.5}$ [erg s$^{-1}$] 
have luminosity dependent systematic error depending on the metallicity calibrator, 
no correlation was seen for H$\alpha$ luminosity $> 10^{36.5}$ [erg s$^{-1}$]. 
Hence, we consider that the correlation between metallicity and SFR 
of H\emissiontype{II} regions would not be strong even if it exists, 
although the existence of the correlation is not robustly ruled out. 

The metallicities of the nearby H\emissiontype{II} regions 
discussed in this section are measured using different metallicity calibration methods 
than the KK04 calibration which we utilize in this study, 
and it is known that different metallicity calibrations 
are often inconsistent with each other \citep{Kewley:2008a}. 
We consider this possible inconsistency as a part 
of the uncertainty of $\sigma_{Z,{\rm int}}$ which we discuss in section~\ref{sec:sigma_Zint}. 

It is broadly agreed that the metallicity decreases 
as the galactocentric radius increases in many galaxies (e.g., \cite{Shields:1978a}). 
However, we note that a scatter around the gradient at each radius is also significant. 
The scatter of metallicity at each galactocentric radius 
can be comparable to (or evan larger than) 
the radial variation (e.g., \cite{Sanders:2012c}), 
and lowest-metallicity H\emissiontype{II} regions in a galaxy 
may reside close to the center of the galaxy. 

It should be noted that the scatter of the metallicities 
in each galaxy may in part result from uncertainties of the metallicity measurements. 
However, similar variation of metallicity is also found in the MW 
using stars with ages of $\sim 10$ Gyr as a tracer of metallicity \citep{Schlesinger:2012a}. 
Furthermore, in the M31 H\emissiontype{II} region sample of \citet{Sanders:2012c}, 
the weak auroral line [O\emissiontype{III}]$\lambda$4363 is detected for four H\emissiontype{II} regions. 
With the electron temperature ($T_e$) method of metallicity measurement 
using the auroral line (e.g., \cite{Garnett:1992a}),
which is not affected by uncertainties of the photoionization models, 
they showed that the four H\emissiontype{II} regions have 12+log$_{10}$(O/H) $\sim$ 8.3--8.4. 
It is generally difficult to detect an auroral line of a high-metallicity H\emissiontype{II} region, 
and it is not surprising that the four H\emissiontype{II} regions 
with the auroral line detections are biased towards low-metallicities. 
However, the $T_e$ method measurements provides 
a confirmation that some H\emissiontype{II} regions in M31 have low-metallicity in reality. 

\subsection{Metallicity of progenitor stars and GRB production efficiency}
\label{sec:eff}

We parameterize the relation between GRB production efficiency 
($\epsilon_{\rm GRB} = R_{\rm GRB}$/SFR) and metallicity of a population of young stars. 
As discussed in section~\ref{sec:Zint}, 
the metallicity of the stellar population 
may be different from that of the host galaxy. 
Here we consider that $\epsilon_{\rm GRB}$ is suppressed 
above a threshold metallicity (O/H)$_{\rm cut}$ 
by a factor of $f_{\rm cont}$ which is not necessarily zero (step function model): 
\begin{eqnarray}
\epsilon_{\rm GRB,s} = \left\{
  \begin{array}{lr}
    \epsilon_{\rm GRB,0} & {\rm O/H \leq (O/H)_{cut}} \\
    f_{\rm cont}\epsilon_{\rm GRB,0} & {\rm O/H > (O/H)_{cut}} 
  \end{array}\right.. \label{eq:eff_step}
\end{eqnarray}
We note that absolute scale of the efficiency (i.e., $\epsilon_{\rm GRB,0}$) 
is marginalized when we consider normalized metallicity PDF of galaxies. 

\begin{figure}
 \begin{center}
  \includegraphics[width=8cm]{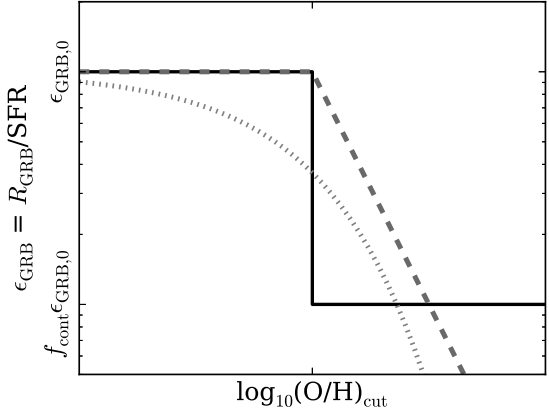}
 \end{center}
 \caption{
   Schematic picture of the step function (solid, eq.~\ref{eq:eff_step}), power-law (dashed, eq.~\ref{eq:eff_pow}), 
   and exponential (dotted, eq.~\ref{eq:eff_exp}) models of $\epsilon_{\rm GRB}$ as functions of metallicity. 
 }\label{fig:eff}
\end{figure}

We also consider the cases with the power-law $\epsilon_{\rm GRB}$ model: 
\begin{eqnarray}
\epsilon_{\rm GRB,p} = \left\{
  \begin{array}{lr}
    \epsilon_{\rm GRB,0} & {\rm O/H \leq (O/H)_{cut}} \\
    \epsilon_{\rm GRB,0}\times[\frac{\rm O/H}{\rm (O/H)_{cut}}]^{\alpha} & {\rm O/H > (O/H)_{cut}} 
  \end{array}\right., \label{eq:eff_pow}
\end{eqnarray}
and with the exponential $\epsilon_{\rm GRB}$ model: 
\begin{equation}
\epsilon_{\rm GRB,e} = \epsilon_{\rm GRB,0}\times{\rm exp}(-[\frac{\rm O/H}{\rm (O/H)_{cut}}]^{\beta}).
\label{eq:eff_exp}
\end{equation}
The schematic picture of the $\epsilon_{\rm GRB}$ models is shown in figure~\ref{fig:eff}

GRB rate of a galaxy with a given set of SFR and 12+log$_{10}$(O/H)$_{\rm gal}$ 
(median metallicity of star forming regions in the galaxy) 
can be obtained by performing the following integration: 
\begin{equation}
R_{\rm GRB,gal} = {\rm SFR} \int \epsilon_{\rm GRB}(Z) \psi(Z) dZ \equiv  {\rm SFR}\ \bar{\epsilon}_{\rm GRB,gal}, 
\label{eq:RGRB}
\end{equation}
where $\psi(Z)$ is the metallicity distribution of star forming regions in the galaxy 
which is represented by a log-normal distribution with a median value 
that corresponds to 12+log$_{10}$(O/H)$_{\rm gal}$ 
and a log scale scatter $\sigma_{Z,{\rm int}}$ (see section~\ref{sec:Zint}). 

\section{Results}
\label{sec:result}

Assuming $\phi(M_\star)$, the $M_\star$-SFR relation, and the $M_\star$-$Z$ relation, 
we compute number density of galaxies in the parameter space of $M_\star$, SFR, $Z$: 
$\varrho(M_\star, {\rm SFR}, Z)$ [Mpc$^{-3}$dex$^{-3}$]. 
The expected metallicity distribution of GRB host galaxies can be derived 
integrating $\varrho(M_\star, {\rm SFR}, Z)$ over $M_\star$ and SFR 
weighted with $R_{\rm GRB,gal}$ (equation~\ref{eq:RGRB}). 

Each $\epsilon_{\rm GRB}$ model discussed in section~\ref{sec:eff} 
has two free parameters: [(O/H)$_{\rm cut}$, $f_{\rm cont}$], 
[(O/H)$_{\rm cut}$, $\alpha$], or [(O/H)$_{\rm cut}$, $\beta$]. 
The parameter (O/H)$_{\rm cut}$ determines the cutoff metallicity 
above which the GRB production efficiency is suppressed, 
and $f_{\rm cont}$, $\alpha$, or $\beta$ determines the sharpness of the cutoff. 
We search for parameter sets that reproduces 
the observed metallicity distribution of the low-redshift sample, 
through the parameter ranges $7.0<12+{\rm log_{10}(O/H)_{cut}}<9.5$, 
$-8.0<{\rm log_{10}}f_{\rm cont}<0.0$, $-8.0<\alpha<0.0$, and $0.0<\beta<8.0$ 
with intervals of $\Delta{\rm log_{10}(O/H)_{cut}}=0.01$, 
and $\Delta{\rm log_{10}}f_{\rm cont} = \Delta\alpha = \Delta\beta = 0.2$. 

\begin{figure*}
 \begin{center}
  \includegraphics[width=15cm]{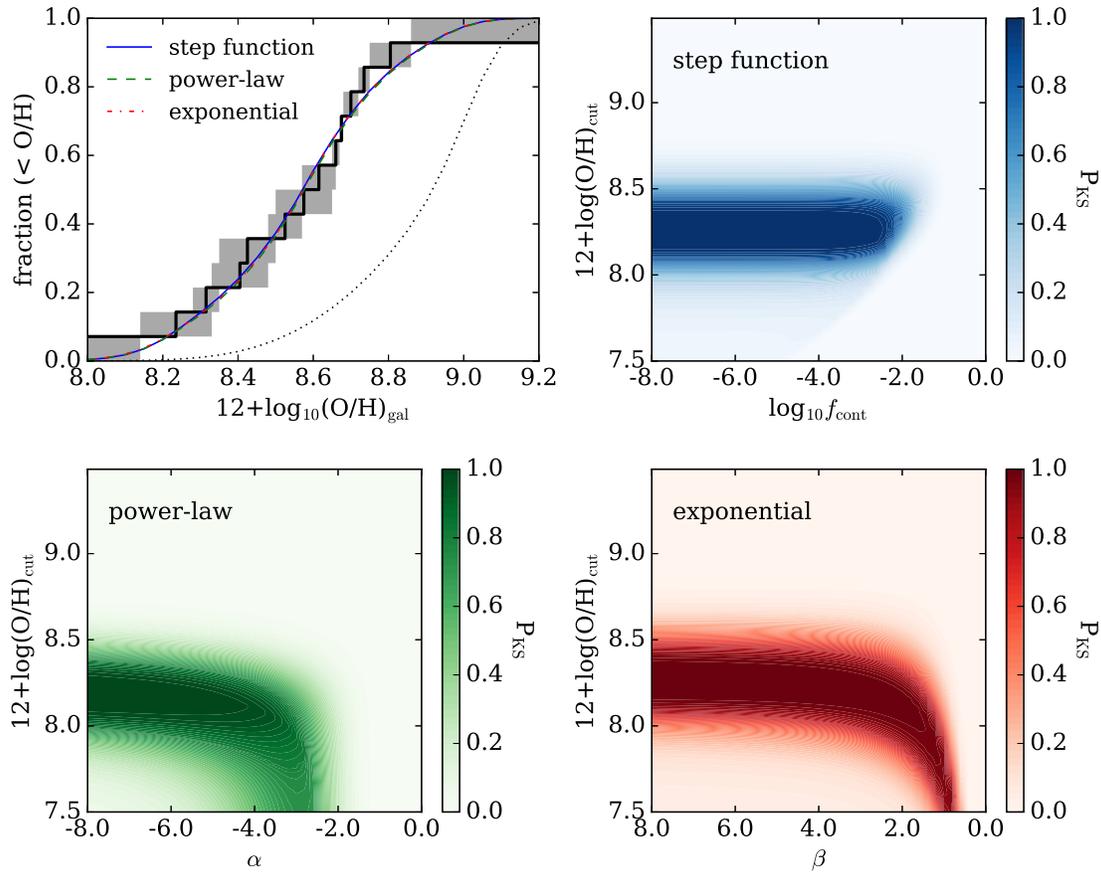}
 \end{center}
 \caption{
   {\it Upper left panel}: the cumulative metallicity distribution 
   of the low-redshift sample of GRB host galaxies a $z < 0.4$ (histogram), 
   and the best fit model distributions with the three different $\epsilon_{\rm GRB}$ models 
   (solid, dashed, and dot-dashed lines). 
   The SFR weighted metallicity distribution is plotted together (dotted line). 
   The gray shaded region associated with the histogram 
   indicates the error due to the uncertain metallicities 
   of the host galaxies of GRB 061021, 111225A, and 090417B. 
   {\it Upper right, lower left, and lower right panels}: the $P_{KS}$ likelihood 
   distribution as a function of the $\epsilon_{\rm GRB}$ model parameters 
   for the step function, power-law, and exponential models, respectively. 
 }\label{fig:result}
\end{figure*}

The goodness of fit is determined using the Kolmogorov-Smirnov (KS) test 
against the ``median'' distribution shown in the upper left panel of figure~\ref{fig:result}. 
Although the metallicity of the GRB 111225A host galaxy is not measured 
and those of the host galaxies of GRB 061021 and 090417B 
are not precisely constrained (see section~\ref{sec:metal}), 
the uncertainties of these metallicities do not significantly 
affect the overall metallicity distribution of GRB host galaxies. 
The resulting parameters from the fitting are not significantly 
changed within the range of this uncertainty. 

The bestfit model distributions are plotted in the upper left panel of figure~\ref{fig:result} 
together with the metallicity distribution of the low-redshift sample. 
The best fitting metallicity distributions with the three different 
$\epsilon_{\rm GRB}$ models are very similar to each other. 
The SFR weighted metallicity distribution predicted from the same $\phi(M_\star)$, 
the $M_\star$-SFR relation, and the $M_\star$-$Z$ relation, 
which the GRB host galaxies would follow if $\epsilon_{\rm GRB}$ 
is not dependent on metallicity, is also shown together. 

The acceptable range of the parameters are shown 
as contour maps of the K-S test probability $P_{\rm KS}$ in figure~\ref{fig:result}. 
The parameter ranges that the $P_{\rm KS} > 0.32$ are:  
\begin{description}
\item[step function: ] 
12+log$_{10}$(O/H)$_{\rm cut} = 8.28^{+0.28}_{-0.31}$,  $f_{\rm cont} < 10^{-1.8}$
\item[power-law: ]
12+log$_{10}$(O/H)$_{\rm cut} < 8.49$,  $\alpha < -2.2$
\item[exponential: ]
12+log$_{10}$(O/H)$_{\rm cut} < 8.57$,  $\beta > 0.6$
\end{description}
The degeneracy between the two parameters is seen 
in the cases of the power-law and exponential models. 
The best fit 12+log$_{10}$(O/H)$_{\rm cut}$ is $\sim$ 8.3, 
and the sharper cutoff reproduces the observations better in any of the models. 
This means that majority of low-redshift GRB host galaxies have higher metallicity 
than 12+log$_{10}$(O/H)$_{\rm cut}$ above which GRBs cannot be produced. 
In other words, most low-redshift GRBs take place in star forming regions 
whose local metallicity is much lower than the representative value of their host galaxies. 

The model distribution with the step function model of $\epsilon_{\rm GRB}$ 
does not significantly depend on $f_{\rm cont}$ when log$_{10}f_{\rm cont} \lesssim -3$, 
and the distribution with $f_{\rm cont} = 0.0$ is quite similar 
to the best fit distribution shown in the upper left panel of figure~\ref{fig:result}. 
We also note that, in the limit of $\alpha = -\infty$ ($\beta = \infty$), 
the power-law (exponential) model of $\epsilon_{\rm GRB}$ 
is identical to the step function model with $f_{\rm cont} = 0.0$. 
In the following, we only consider the step function model 
of $\epsilon_{\rm GRB}$ with $f_{\rm cont} = 0.0$ to examine 
the predicted properties of low-redshift GRB host galaxies with our models. 

\begin{figure}
 \begin{center}
  \includegraphics[width=8cm]{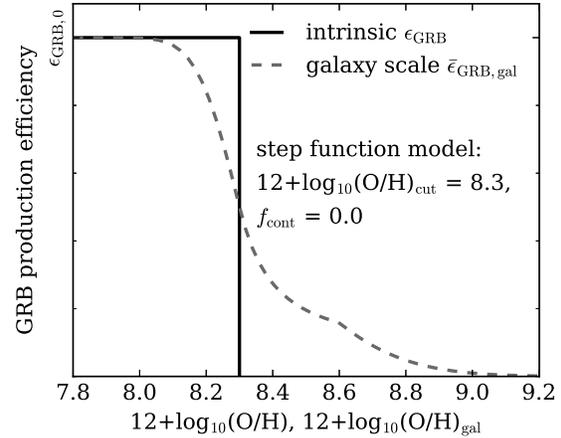}
 \end{center}
 \caption{
   The GRB production efficiency ($\epsilon_{\rm GRB}$) of a stellar population 
   with a single metallicity [solid line, the step function model 
   with 12+log$_{10}$(O/H)$_{\rm cut}$ = 8.3 and $f_{\rm cont} = 0.0$], 
   and the efficiency convolved over galaxy scale 
   taking the internal variation of metallicity within each galaxy 
   into account (dashed line, see eq.~\ref{eq:RGRB}).  
 }\label{fig:effgal}
\end{figure}

In figure~\ref{fig:effgal}, we compare the step function model 
of $\epsilon_{\rm GRB}$ [12+log$_{10}$(O/H)$_{\rm cut}$ = 8.3, $f_{\rm cont} = 0.0$] 
with that convolved over galaxy scale ($\bar{\epsilon}_{\rm GRB,gal}$) 
as defined in equation~\ref{eq:RGRB}. 
$\bar{\epsilon}_{\rm GRB,gal}$ steeply declines 
with increasing metallicity at around 12+log$_{10}$(O/H)$_{\rm cut}$. 
However, galaxies with 12+log$_{10}$(O/H)$_{\rm gal} >$ 12+log$_{10}$(O/H)$_{\rm cut}$ 
have $\bar{\epsilon}_{\rm GRB,gal} > 0.0$ due to the internal variation of metallicity, 
and their contribution to the cosmic GRB rate density is significant 
because they play a dominant role in the cosmic star formation at low-redshifts. 
The steep decline of $\bar{\epsilon}_{\rm GRB,gal}$ 
at 12+log$_{10}$(O/H)$_{\rm gal} \sim 8.2$ is also reported by \citet{Graham:2015a} 
based on a GRB host galaxy sample collected from a wider range of redshifts 
where metallicity measurements of GRB host galaxies are incomplete. 
We note that $\bar{\epsilon}_{\rm GRB,gal}(Z)$ likely evolve with redshift 
because it depends on the properties of galaxies (e.g., $\sigma_{Z,{\rm int}}$), 
unlike $\epsilon_{\rm GRB}(Z)$ which would be determined by stellar physics. 

In some cases, GRB positions are spatially resolved 
from the center (or brightest part) of the host galaxies (e.g., \cite{Modjaz:2008a}). 
\citet{Levesque:2011a} has shown that the metallicities at the GRB positions  
are systematically lower than that of their host galaxies by $\sim 0.1$ dex, 
but not necessarily as low as suggested by our results. 
\citet{Niino:2015a} pointed out that the actual metallicity of a GRB explosion site 
cannot be obtained with a spatial resolution $\gtrsim 1$ kpc 
which is typical of the observations of GRB positions at $z > 0.1$, 
because the length scale of the ISM metallicity variation 
is $< 1$ kpc in nearby galaxies (e.g., \cite{Sanders:2012c}). 
Our results suggest that the actual metallicity of the GRB explosion sites (or the progenitor star itself)
which is buried within the spatial resolution is significantly lower 
than the kpc scale metallicity currently observed, by up to $\sim$ 0.5 dex. 

\begin{figure}
 \begin{center}
  \includegraphics[width=8cm]{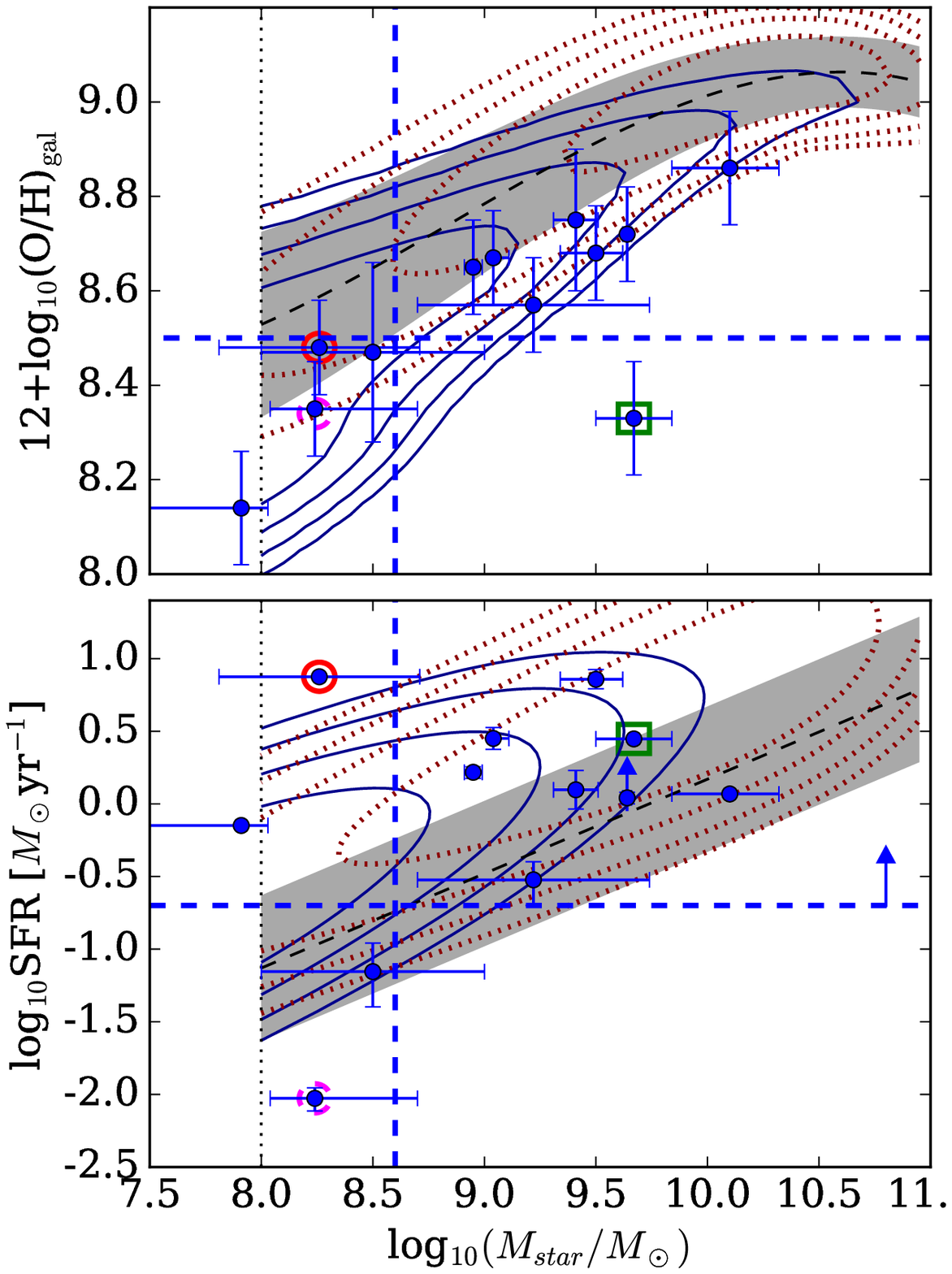}
 \end{center}
 \caption{
    The predicted and observed distribution of GRB host galaxies 
    on the parameter planes of $M_\star$ vs. $Z$ ({\it upper panel}), 
    and $M_\star$ vs. SFR ({\it lower panel}). 
    The solid and dotted contours represent the predicted distributions 
    weighted with the $R_{\rm GRB,gal}$ and SFR, respectively. 
    The contours are drawn at a logarithmic interval 
    of 0.25 dex from the peaks of the distributions. 
    The vertical dashed lines indicate naively estimated $M_\star$ of the host galaxy of GRB 111225A 
    for which 12+log$_{10}$(O/H)$_{\rm gal}$ and SFR is not measured.  
    The horizontal dashed lines represent 12+log$_{10}$(O/H)$_{\rm gal}$ 
    and the lower-limit SFR of the host galaxy of GRB 120714B whose $M_\star$ is not known. 
    The dashed lines with gray shaded regions are 
    the $M_\star$-$Z$ relation \citep{Kewley:2008a} 
    and the $M_\star$-SFR relation \citep{Salim:2007a}. 
    The shaded regions represent 1$\sigma$-scatter of the relations. 
    The following host galaxies are marked with specific symbols: 
    GRB 011121 (square, $M_\star$-$Z$ outlier), 
    031203 (solid circle, possible AGN, $M_\star$-SFR outlier), 
    060614 (dashed circle, $M_\star$-SFR outlier). 
    The vertical dotted line indicates the lower-limit of the $M_\star$ range 
    which we consider when we predict the metallicity distribution of GRB host galaxies.  
 }\label{fig:2Ddist}
\end{figure}

The $R_{\rm GRB,gal}$ weighted distribution of galaxies 
on the $M_\star$-$Z$ and $M_\star$-SFR parameter planes, 
predicted using the step function model of $\epsilon_{\rm GRB}$  
with 12+log$_{10}$(O/H)$_{\rm cut}$ = 8.3 and $f_{\rm cont} = 0.0$, 
is plotted as contours in figure~\ref{fig:2Ddist}. 
The SFR weighted model distribution, the assumed $M_\star$-$Z$ relation \citep{Kewley:2008a}, 
the $M_\star$-SFR relation \citep{Salim:2007a}, and the observed properties 
of the low-redshift sample (table~\ref{tab:metal}~\&~\ref{tab:msfr}) are plotted together. 

The observed GRB host galaxies construct a sequence 
which is offset from the $M_\star$-$Z$ relation by 0.1--0.2 dex towards lower-metallicities, 
as previously noted by \citet{Levesque:2010a} with a smaller incomplete sample, 
except for one remarkable outlier: the GRB 011121 host galaxy, 
whose N2 and $R_{23}$ metallicities do not agree well with each other. 
The $R_{\rm GRB,gal}$ weighted model distribution also shows similar offset on the $M_\star$-$Z$ plane due 
to the higher $\bar{\epsilon}_{\rm GRB,gal}$ in galaxies with lower 12+log$_{10}$(O/H)$_{\rm gal}$. 
Although the offset of the observed GRB host galaxies 
may be slightly larger than that of the model distribution, 
this possible contradiction can be resolved by 
taking the SFR-$Z$ correlation into account
as we discuss in section~\ref{sec:MSFRZ}. 

In the $M_\star$-SFR plane, the predicted galaxy distributions 
with the SFR and $R_{\rm GRB,gal}$ weightings are offset from the $M_\star$-SFR relation 
by $\sim$ 0.5 dex ($\sim \sigma_{\rm MSFR}$, as generally expected 
when the scatter of a relation is represented by a log-normal distribution). 
The observed GRB host galaxies are broadly consistent with the predicted offset, 
although the SFR scatter of the host galaxies 
with log$_{10}M_\star/M_\odot < 9.0$ is large. 

The host galaxy of GRB 031203, which might have an AGN, 
agrees well with the sequence of other GRB host galaxies in the $M_\star$-$Z$ plane, 
while it is largely offset towards higher-SFR in the $M_\star$-SFR plane. 
The outliner in the $M_\star$-$Z$ plane, the GRB 011121 host galaxy, 
does not show any peculiarity in the $M_\star$-SFR plane. 
The host galaxy with the lowest SFR is that of GRB 060614 which is not associated with a SN. 
The GRB 060614 host galaxy does not show any peculiarity in the $M_\star$-$Z$ plane. 

\section{Discussion}
\label{sec:discussion}

\subsection{Uncertainties in the models of galaxies}
\label{sec:model_check}

The model predictions of the metallicity distribution 
depend not only on the $\epsilon_{\rm GRB}$ model, 
but also on the underlying assumptions of the properties of galaxies. 
In this section, we examine how our results are affected 
by some galaxy properties which is not well understood, 
namely the SFR-$Z$ correlation and the variation of metallicity within a galaxy. 

\subsubsection{The $M_\star$-SFR-$Z$ relation of galaxies}
\label{sec:MSFRZ}

To investigate the effect of the SFR-$Z$ correlation, 
we consider the $M_\star$-SFR-$Z$ relation by \citet{Mannucci:2011a}, 
which is an extension of the relation by \citet{Mannucci:2010a} towards lower-$M_\star$. 
However, the metallicity calibration method used 
in \citet{Mannucci:2011a} is different from that used in this study (KK04). 
To examine the effect of the SFR-$Z$ correlation in a consistent metallicity scale, 
we compute the $\frac{\partial {\rm log_{10}(O/H)}}{\partial {\rm log_{10}SFR}}$ gradient  
of the $M_\star$-SFR-$Z$ relation around the galaxy main-sequence at each $M_\star$. 
And assume the gradient without changing the $M_\star$-$Z$ relation \citep{Kewley:2008a}, 
which has been used in the previous sections. 
 
It should be noted that the scatter of the $M_\star$-$Z$ relation 
at each $M_\star$ may partly result from the SFR-$Z$ correlation. 
\citet{Mannucci:2010a} showed that the metallicity scatter around the $M_\star$-SFR-$Z$ relation 
is smaller by 50\% than that of the $M_\star$-$Z$ relation. 
Thus we assume the scatter of the $M_\star$-SFR-$Z$ relation
is 50\% of that of the $M_\star$-$Z$ relation defined in equation~\ref{eq:sigmaMZ}. 

\begin{figure}
 \begin{center}
  \includegraphics[width=8cm]{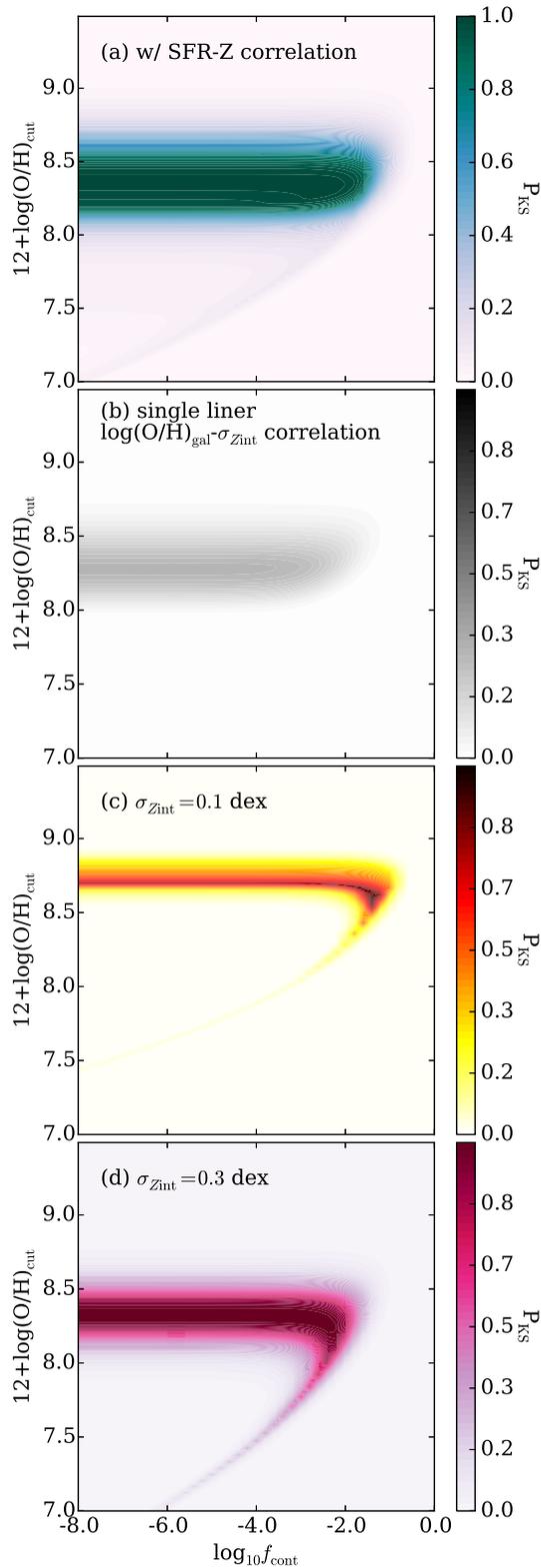}
 \end{center}
 \caption{
   Same as the upper right panel of figure~\ref{fig:result}, 
   but with different assumptions of the properties of galaxies. 
   The step function model of $\epsilon_{\rm GRB}$ is utilized. 
   {\it (a)}: the correlation between SFR and log$_{10}$(O/H) is incorporated. 
   {\it (b)}: $\sigma_{Z,{\rm int}}$ is linearly correlated with 12+log$_{10}$(O/H)$_{\rm gal}$. 
   {\it (c) and (d)}: $\sigma_{Z,{\rm int}}$ is a constant value (0.1 and 0.3 dex, respectively). 
 }\label{fig:gmodel_pks}
\end{figure}

\begin{figure}
 \begin{center}
  \includegraphics[width=8cm]{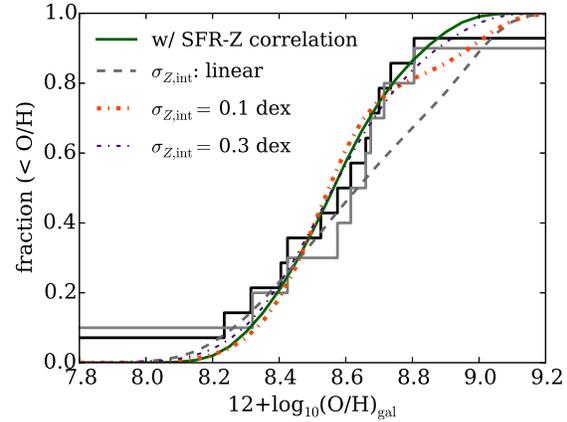}
 \end{center}
 \caption{
   Same as the upper left panel of figure~\ref{fig:result}, 
   but with different assumptions of the properties of galaxies. 
   The step function model of $\epsilon_{\rm GRB}$ is utilized. 
   The plotted models are the same as those investigated in figure~\ref{fig:gmodel_pks}. 
   The black histogram is the same as that in the upper left panel of figure~\ref{fig:result}, 
   while the gray histogram represents the metallicity distribution 
   without the host galaxies of GRB 011121, 031203, 060505, and 060614 
   whose burst classifications or metallicity measurements are uncertain. 
 }\label{fig:gmodel_dist}
\end{figure}

The resulting $P_{KS}$ contour map and the bestfit metallicity distribution 
using the step function $\epsilon_{\rm GRB}$ model 
are shown in figure~\ref{fig:gmodel_pks} (top panel) and figure~\ref{fig:gmodel_dist}, respectively. 
Both the best fit parameters and the goodness of the fit 
are not significantly different from the case without the SFR-$Z$ correlation. 
This situation is similar with the power-law and exponential models of $\epsilon_{\rm GRB}$. 

The small effect of the SFR-$Z$ correlation on the metallicity distribution is not surprising. 
The SFR-$Z$ gradient around the galaxy main-sequence is 
$\frac{\partial {\rm log_{10}(O/H)}}{\partial {\rm log_{10}SFR}} = -0.16$ 
in the low-$M_\star$ range (log$_{10}M_\star/M_\odot \lesssim 9$) 
and shallower in higher-$M_\star$ ranges. 
Given that the GRB host galaxies have systematically higher-SFR 
than the main-sequence by $\sim$ 0.5 dex (section~\ref{sec:result}), 
the expected effect of the SFR-$Z$ correlation is $\lesssim 0.08$, 
which is smaller than the uncertainties of the metallicity measurements. 

\begin{figure}
 \begin{center}
  \includegraphics[width=8cm]{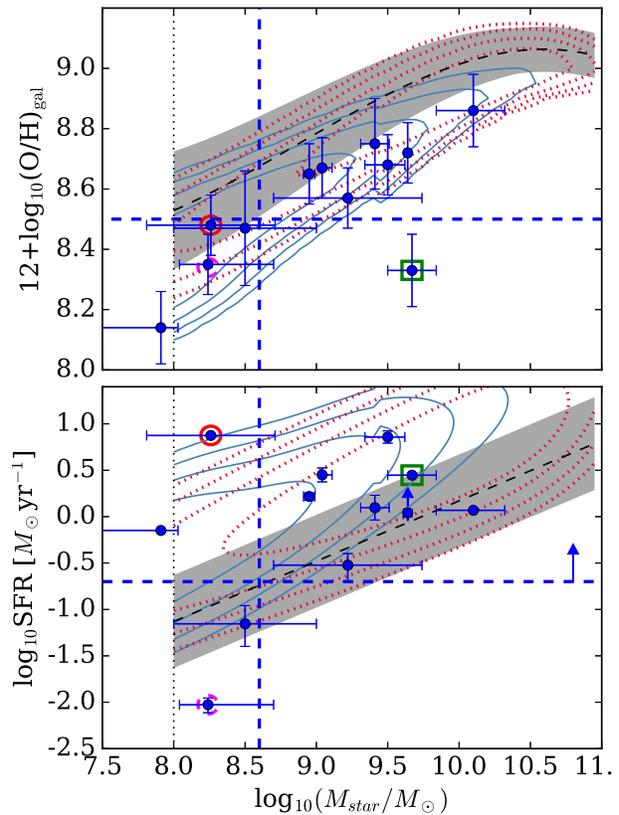}
 \end{center}
 \caption{
    Same as figure~\ref{fig:2Ddist}, but with the  SFR-$Z$ correlation. 
    The anomalies in the contour lines at log$_{10}M_\star/M_\odot \sim 9.5$ 
    result from a discontinuity of the $M_\star$-SFR-$Z$ relation by \citet{Mannucci:2011a}. 
 }\label{fig:2Ddist_FMR}
\end{figure}

The SFR and $R_{\rm GRB,gal}$ weighted model distributions of galaxies 
on the $M_\star$-$Z$ and $M_\star$-SFR planes, 
predicted assuming the SFR-$Z$ gradient are shown in figure~\ref{fig:2Ddist_FMR}. 
The $\epsilon_{\rm GRB}$ model and the parameters are the same as in figure~\ref{fig:2Ddist}. 
The SFR weighted galaxy distribution (i.e., no metallicity effect) 
on the $M_\star$-$Z$ plane is also shifted 
towards lower-metallicities due to the SFR-$Z$ correlation, 
but the shift is smaller than that of the $R_{\rm GRB,gal}$ weighted distribution. 
Furthermore, the GRB host galaxies have systematically lower-$M_\star$ 
than the typical $M_\star$ of the SFR weighted population (log$_{10}M_\star/M_\odot \sim 10$), 
which cannot be explained solely by the effect 
of SFR-$Z$ correlation (e.g., \cite{Campisi:2011a}). 

Although the SFR-$Z$ correlation does not significantly affect 
the overall metallicity distribution of the GRB host galaxies as mentioned above, 
the predicted number of GRB host galaxies above the $M_\star$-$Z$ relation 
is reduced (figure~\ref{fig:2Ddist_FMR}, top panel). 
This improves the consistency between the predicted 
and observed distributions of GRB host galaxies
on the $M_\star$-$Z$ plane compared with the case 
without the SFR-$Z$ correlation (figure~\ref{fig:2Ddist}). 

\subsubsection{The internal $\sigma_{Z,{\rm int}}$ of galaxies}
\label{sec:sigma_Zint}
 
As mentioned in section~\ref{sec:Zint}, 
the detailed investigation of the internal ISM metallicity variation 
is performed only for a small number of nearby galaxies. 
Therefore, $\sigma_{Z,{\rm int}}$ of general star forming galaxies 
and its correlation with other galaxy properties are highly uncertain. 
Given that the nearby high-metallicity galaxies, namely the MW and M31, 
have higher $M_\star$ than the GRB host galaxies in our sample, 
it is possible that ISM in the GRB host galaxies have different properties to those in the nearby galaxies. 

To investigate how our results depend on the assumed $\sigma_{Z,{\rm int}}$, 
we perform the parameter fittings assuming 
other $\sigma_{Z,{\rm int}}$ formulations than the baseline model. 
Here we consider a single linear relation 
between 12+log$_{10}$(O/H)$_{\rm gal}$ and $\sigma_{Z,{\rm int}}$, 
and also constant $\sigma_{Z,{\rm int}}$ of 0.1, 0.2, and 0.3 dex (figure~\ref{fig:Zint}, bottom panel). 
However, it should be noted that the constant $\sigma_{Z,{\rm int}}$ is disfavored by observations. 

The $P_{KS}$ contour maps and the predicted metallicity distributions 
are shown in figure~\ref{fig:gmodel_pks} and figure~\ref{fig:gmodel_dist} 
for the cases with the single linear relation, 
and the constant $\sigma_{Z,{\rm int}}$ of 0.1 and 0.3 dex. 
The single linear relation between 12+log$_{10}$(O/H)$_{\rm gal}$ and $\sigma_{Z,{\rm int}}$ 
overpredicts the contribution of high-metallicity galaxies 
to the GRB production even with the best fit parameters, 
due to the large $\sigma_{Z,{\rm int}}$ of galaxies with 12+log$_{10}$(O/H)$_{\rm gal} > 8.8$. 
However, the statistical significance of the overprediction is low ($P_{\rm KS} = 0.29$). 
The best fit parameters are similar to the case with the baseline model. 

When $\sigma_{Z,{\rm int}} = 0.1$ dex independently of 12+log$_{10}$(O/H)$_{\rm gal}$, 
the preferred cutoff metallicity is 
12+log$_{10}$(O/H)$_{\rm cut} = 8.60^{+0.20}_{-0.25}\ (\sim Z_\odot)$.
This result is consistent with the previous studies 
without the internal metallicity variation being explicitly treated 
in which the cutoff of the GRB efficiency at 0.5--1$Z_\odot$ is suggested 
(e.g., \cite{Wolf:2007a, Kocevski:2009a, Perley:2016b, Vergani:2015a, Japelj:2016a}). 
Even in this case, high $\epsilon_{\rm GRB}$ 
in high-metallicity stellar population is disfavored ($f_{\rm cont} < 0.1$). 

When $\sigma_{Z,{\rm int}} = 0.3$ dex, 
the cutoff metallicity is 12+log$_{10}$(O/H)$_{\rm cut} = 8.31^{+0.23}_{-0.79}$, 
which is close to the results with the baseline model. 
The results with the the constant $\sigma_{Z,{\rm int}} =$ 0.2 dex 
are similar with those with 0.1, and 0.3 dex, 
but with 12+log$_{10}$(O/H)$_{\rm gal} = 8.49^{+0.24}_{-0.46}$. 

\subsection{Possible subpopulations}
\label{sec:subpop}

Some of the low-redshift GRBs are less energetic than typical GRBs 
that are observed up to higher-redshifts (hereafter classical GRBs), 
and possibly are different kind of phenomena.  
These bursts, so called low-luminosity GRBs, have isotropic equivalent gamma-ray energy 
$E_{\rm iso} \lesssim 10^{50}$ [erg] (e.g., \cite{Bromberg:2011a, Nakar:2015a}). 
\citet{Stanek:2006a} have shown that $E_{\rm iso}$ of a GRB 
might be dependent on the metallicity of its host galaxy 
based on observations of the host galaxies of 5 GRBs/XRFs at redshifts $\leq 0.25$, 
while later studies \citep{Levesque:2010g, Japelj:2016a} have shown that there is no significant correlation 
between them with a larger sample from wider range of redshifts (up to $z = 1$). 

\begin{figure*}
 \begin{center}
  \includegraphics[width=15cm]{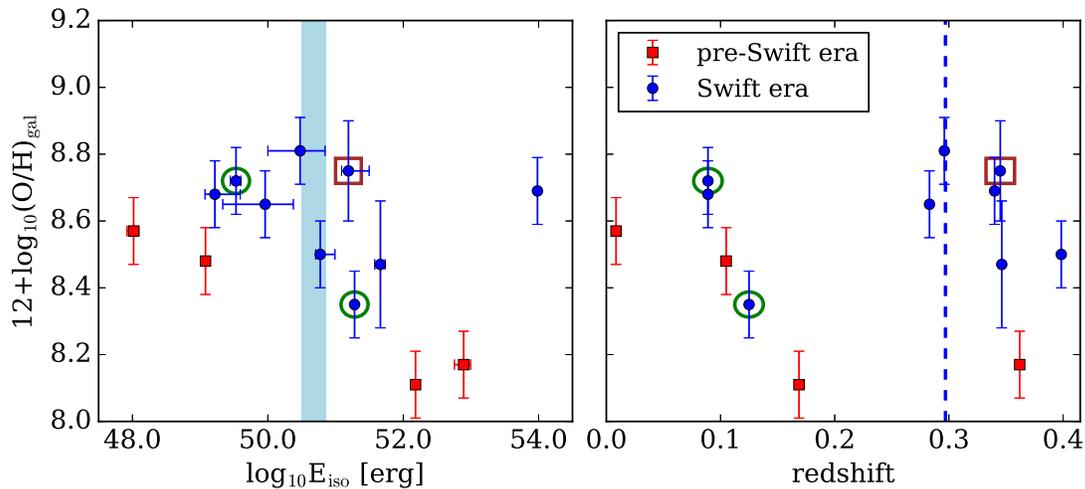}
 \end{center}
 \caption{
    The $E_{\rm iso}$ vs. host metallicity plot ({\it left panel}) 
    and the redshift vs. host metallicity plot ({\it right panel}) of the low-redshift GRBs. 
    GRBs that occurred before and after the launch of the {\it Swift} satellite (2004) 
    are shown with different symbols (red squares and blue circles, respectively). 
    The data points surrounded by (green) circles represent GRBs 
    without SN associations (GRB 060505 and 060614). 
    The data point surrounded by a (brown) square represents ``dark'' GRB 090417B. 
    The shaded region ({\it left panel}) and the of vertical dashed line ({\it right panel})
    indicate $E_{\rm iso}$ and redshift of the host galaxy of GRB 111225A 
    for which 12+log$_{10}$(O/H)$_{\rm gal}$ is not known.  
 }\label{fig:Eiso_z}
\end{figure*}

Because low-luminosity GRBs can be detected only at low-redshifts $\lesssim 0.3$, 
the comparison of the host galaxies and low-luminosity GRBs 
and with those of classical GRBs at $z \sim 1$ might be 
affected by evolution of galaxy populations over the redshits. 
Thus it is interesting to revisit the correlation between $E_{\rm iso}$ and host metallicity 
with our sample limited to low-redshifts but much larger than that of \citet{Stanek:2006a}. 
$E_{\rm iso}$ of the low-redshift GRBs and 12+log$_{10}$(O/H)$_{\rm gal}$ 
are shown in the left panel of figure~\ref{fig:Eiso_z}. 

Although the GRBs in the pre-{\it Swift} era (until 2004) 
shows a trend that GRBs with larger $E_{\rm iso}$ occurr 
in lower-metallicity galaxies as reported in \citet{Stanek:2006a}, 
we do not find significant correlation between $E_{\rm iso}$ and 12+log$_{10}$(O/H)$_{\rm gal}$ 
when GRBs in the {\it Swift} era are included (correlation coefficient is -0.36). 

GRB 060505 and 060614 are widely known as GRBs which are not associated 
with SNe \citep{Gehrels:2006a, Fynbo:2006a, Della-Valle:2006a, Gal-Yam:2006a}, 
and it is possible that they are actually short GRBs 
that do not result from core-collapse of massive stars 
despite their long duration $> 2$ sec (e.g., \cite{Zhang:2009a}). 
These bursts are marked with circles in figure~\ref{fig:Eiso_z}. 
The host metallicities and $E_{\rm iso}$ of GRB 060505 and 060614 
are typical of low-redshift GRBs in our sample. 
The metallicity distributions of the GRB host galaxies excluding 
GRB hosts with possible systematic error in their metallicity measurements 
or with uncertain burst classifications (GRB 011121, 031203, 060505, and 060614) 
is shown in figure~\ref{fig:gmodel_dist}. 
The metallicity distribution without these peculiar objects 
does not significantly differ from the distribution of the whole sample. 

\subsection{Sampling biases}
\label{sec:bias}

In this study, we have presented the overall metallicity distribution 
of the host galaxies of all GRBs known at $z < 0.4$.  
However, the fraction of GRBs with spectroscopic redshifts 
are $\sim$ 30\% even in recent observations (e.g., \cite{Perley:2016a}), 
and hence we may suffer from the redshift determination bias 
by collecting GRBs with known redshifts. 

\subsubsection{Optically ``dark'' GRBs}
\label{sec:darkGRB}

One possible source of the redshift determination bias 
is ``dark'' GRBs (e.g., \cite{Jakobsson:2004a, Greiner:2011a}). 
A ``dark'' GRB is a GRB with a fainter optical afterglow than expected from its X-ray spectrum. 
Roughly 1/3 of GRBs are ``dark'', 
and most of them ($\sim 80$\%) result from dust extinction, 
while some may result from absorption 
by neutral hydrogen in inter-galactic medium at high-redshifts 
\citep{Cenko:2009a, Zheng:2009a, Greiner:2011a}. 
Redshift determination of GRBs largely relies on optical spectroscopy of afterglows, 
and it is difficult to determine redshifts of ``dark'' GRBs. 
Thus a sample of GRB host galaxy with known redshifts may be biased 
against dusty galaxies which may tend to have higher-metallicity. 

However, at low-redshifts, detection of a GRB host galaxy 
is not quite challenging in most of the cases. 
Assuming the stellar-mass to luminosity ratio derived by \citet{Kauffmann:2003b}, 
a galaxy with $M_\star = 10^9M_\odot$, 
which is typical of the low-redshift GRB host galaxies, 
would be observed with $r-$band magnitude $\sim 24$ at $z = 0.3$. 
Once the host galaxy of a GRB is detected, 
X-ray afterglow localization by {\it Swift}/XRT with typical precision of a few arcseconds, 
which is available for $\sim 80$\% of the {\it Swift}/BAT-detected GRBs, 
can associate the GRB to the host galaxy with a certain level of confidence \citep{Perley:2013a}. 
And the redshift can be determined using emission lines of the host galaxy. 

In reality, many redshifts of the low-redshift GRBs are determined 
by the emission lines of the host galaxies (table~\ref{tab:GRBs}), 
including that of a ``dark'' GRB 090417B (marked with a square in figure~\ref{fig:Eiso_z}), 
while redshift determination by host emission lines is rarely available at higher redshifts. 
The fraction of the GRBs whose redshifts are determined 
only by the host emission lines in the low-redshift sample (8/14) is naively comparable 
to the fraction of GRBs without known redshifts in the {\it Swift}/BAT-detected GRBs. 
Thus the bias against GRBs in dusty environment 
is not necessarily effective at low-redshifts. 

\subsubsection{GRBs in faint host galaxies}
\label{sec:fainthost}

It is possible that GRBs whose redshifts are determined 
with the emission lines of their host galaxies are biased 
against GRBs that occurred in galaxies with weak emission line fluxes. 
Emission line luminosities of a galaxy is largely dependent on SFR in the galaxy, 
which correlates with $M_\star$ and 12+log$_{10}$(O/H)$_{\rm gal}$, 
in the sense that galaxies with higher $M_\star$ 
and 12+log$_{10}$(O/H)$_{\rm gal}$ have more luminous emission lines. 
Hence GRBs that occurred in very low-metallicity host galaxies 
might be systematically missed in the low-redshift sample, 
especially at higher-end of the redshift range 
where detection of weak emission lines are difficult. 
However, such redshift dependent metallicity bias is not seen in our sample 
(the right panel of figure~\ref{fig:Eiso_z}, correlation coefficient is 0.007). 

It is also interesting to compare the expected rate of low-redshift GRBs 
to the actual finding rate to evaluate the success rate of identifying low-redshift GRBs. 
Based on the largest unbiased GRB survey to date (SHOALS), 
\citet{Perley:2016a} estimated the event rate density 
of bright GRBs with $E_{\rm iso} > 10^{51}$ [erg]
at $z < 0.5$ to be $0.2^{+0.3}_{-0.1}$ [yr$^{-1}$Gpc$^{-3}$]. 
Considering the field of view of {\it Swift}/BAT (1.4 steradian) 
and the operation time until the end of March 2014 (9.3 years since December 2004), 
the expected number of bright GRBs detected by {\it Swift}/BAT 
at $z < 0.4$ in this time period is $3.2^{+4.9}_{-1.6}$. 
On the other hand, the number of {\it Swift}/BAT-detected GRBs 
with $E_{\rm iso} > 10^{51}$ [erg] in our sample is 4 (figure~\ref{fig:Eiso_z}), 
which suggests that the success rate of redshift determination 
is higher at low-redshifts compared with that of the overall population of GRBs, 
although the estimation of the GRB rate density is still uncertain. 
The faint end of the GRB luminosity function 
is still controversial affected 
by sample selections and model assumptions 
(e.g., \cite{Schmidt:2001a, Wanderman:2010a, Salvaterra:2012a}). 
Hence it is difficult to estimate the actual rate 
and evaluate the redshift success rate of low-luminosity GRBs. 

On the other hand, it is also true that redshift determination of a GRB has failed 
(partly) due to the weak emission line of the host galaxy. 
As mentioned in section~\ref{sec:obs:111225a}, the redshift of GRB 111225A 
has not been determined until \citet{Thone:2014b} reports 
detections of weak emission lines over the afterglow 
together with a marginal absorption line using an updated analysis tool. 
The failure of the redshift determination  
in the immediate follow up observations after the burst
suggests that GRB host galaxies with $M_\star \lesssim 10^8M_\odot$ 
might be missed in redshift selected samples, 
given that the stellar mass of the GRB 111225A host galaxy is 
naively $\sim 4\times10^8M_\odot$ (section~\ref{sec:host_msfr}). 

\subsubsection{Contamination by foreground galaxies}
\label{sec:contamination}

Another interesting case is the identification of the host galaxy of GRB 020819B. 
GRB 020819B has been considered to be a low-redshift burst at $z=0.41$
since radio afterglow observations by \citet{Jakobsson:2005b} 
localized the burst on a ``blob'' which resides 
$\sim$ \timeform{3''} from a spiral galaxy at $z=0.41$, 
and further observations of the host galaxy by \citet{Levesque:2010b} 
suggested that the ``blob'' also resides at $z=0.41$. 
However, recent extensive observations of the host galaxy by \citet{Perley:2016c} 
showed that the ``blob'' is actually a dusty starforming galaxy at $z=1.96$, 
and the low-redshift spiral galaxy is possibly a foreground object unrelated to the GRB.  

Hence it is possible that a few of the low-redshift GRBs 
in our sample whose redshifts are obtained only via their host galaxies 
actually reside in high-redshift galaxies which are too faint to be observed, 
and currently claimed host galaxies at low-redshifts are foreground objects. 
Although most of the low-redshift GRBs are closely located to the center 
of their host galaxies within $\leq$ \timeform{1''} unlike GRB 020819B, 
it is difficult to robustly confirm that a ``low-redshift'' GRB 
is actually at a low-redshift when the GRB does not 
have an afterglow measured redshift or a confirmed SN association 
(GRB 060505, 080517, 060614, 050826, and 090417B, see table~\ref{tab:GRBs}). 
Furthermore, among the low-redshift GRBs without robust redshift confirmations, 
GRB 060505 and 080517 have offsets $\geq$ \timeform{2''} from the center of their host, 
although the large offsets are not surprising given their very low-redshifts $z < 0.1$. 

It should be noted that the originally claimed host galaxy 
of GRB 020819B at $z=0.41$ was a massive-spiral galaxy 
with the highest metallicity in the known GRB host galaxies. 
When a GRB host galaxy sample is contaminated by unrelated forground galaxies, 
it is possible that the contaminants tend to be brighter galaxies 
because they can be found more easily than 
the actual host galaxies which are likely faint. 

Missing redshift informations for GRBs in faint galaxies 
and contamination by bright foreground galaxies 
may bias the metallicity distribution 
of the overall population towards higher-metallicities, 
in contrast to what is expected in the case of 
the bias against ``dark'' GRBs. 

\section{Conclusions}
\label{sec:conclusion}

We have presented the overall metallicity distribution of the host galaxies 
of all GRBs known at $z < 0.4$ that occurred before the end of March 2014, 
including the newly obtained emission line fluxes 
of the host galaxies of GRB 060614, 090417B, and 130427A. 
The low-redshift sample of GRB host galaxies 
with complete metallicity measurements is essential 
to study the nature of GRB progenitor in some aspects: 
1) the high success rate of redshift determinations and host identifications, 
2) the reporting bias can be eliminated 
by completing the metallicity measurements, and
3) a wealth of comparison sample is available. 

We have compared the metallicity distribution of the low-redshift sample 
with the predictions from the empirical formulations 
of the properties of low-redshift galaxies. 
To predict the metallicity distribution, we adopted models of GRB efficiency 
($\epsilon_{\rm GRB}=R_{\rm GRB}/{\rm SFR}$) as functions of metallicity. 
The metallicity of the progenitor stars considered here 
is not necessarily identical to that of its host galaxy. 
Instead, we considered the metallicity variation within each galaxy 
motivated by the observations of H\emissiontype{II} regions in nearby galaxies. 

The three different formulations of $\epsilon_{\rm GRB}$, 
namely the step function, power-law, and exponential models are examined. 
Each of the three $\epsilon_{\rm GRB}$ models has two free parameters 
that can be used to fit the metallicity distribution of the host galaxies. 
In either case, $\epsilon_{\rm GRB}$ function with a sharp cutoff around 
12+log$_{10}$(O/H)$_{\rm cut} \sim$ 8.3 ($\sim 0.4Z_\odot$) 
reproduces the metallicity distribution of the low-redshift sample best. 
This cutoff metallicity is naively consistent with 
those predicted by the stellar evolution models 
($\sim$ 0.1--$0.5Z_\odot$, \cite{Yoon:2005a,Hirschi:2005a,Woosley:2006a}). 
$\epsilon_{\rm GRB}$ models with moderate or weak metallicity dependence 
[$\epsilon_{\rm GRB}(Z_\odot)/\epsilon_{\rm GRB}(0.1Z_\odot) > 0.1$] are disfavored, 
although the statistical significance 
of the current constraints is still low. 

This result is in contrast to the results of some previous studies 
which have suggested the cutoff metallicity $\sim$ 0.5--$1Z_\odot$ 
(e.g., \cite{Wolf:2007a, Kocevski:2009a, Vergani:2015a, Perley:2016b, Japelj:2016a}). 
This is because we take the internal variation 
of metallicity within each galaxy into account. 
Our galaxy models also indicate the cutoff metallicity around 0.5--1$Z_\odot$,
when the internal metallicity variation is assumed to be smaller (0.2--0.1 dex) in any galaxy.
However, the current model of the internal metallicity variation 
relies on the observed metallicity distributions 
of H\emissiontype{II} regions in a small number of local galaxies, 
and the actual internal metallicity variation in GRB host galaxies is hardly known. 
This uncertainty possibly affect our results significantly. 

The effect of the correlation beween SFR and metallicity of galaxies 
on the predicted metallicity distribution is smaller 
than the uncertainty of metallicity measurement. 
However, it may improve the agreement of the predicted and observed distribution 
of the GRB host galaxies on the parameter plane of $M_\star$ versus $Z$. 
The relation between the nature of GRB progenitor stars 
and the observable properties of their host galaxy 
depends on the internal and global properties of galaxies. 
Better understanding on the general population of galaxies at each redshift 
is essential to obtain more robust constraints on the properties of GRB progenitors, 
in addition to gathering larger less biased sample of GRB host galaxies.  

Because we collected GRBs with known redshifts ($< 0.4$) in this study, 
it is possible that the sample of the host galaxies is biased for 
a kind of galaxy population in which GRB redshift determination 
is easy (e.g., galaxies with smaller dust content), 
although the GRB redshift determination is generally easier at lower-redshifts. 
However, the number of low-redshift GRBs in our sample is close to 
the number expected from the latest unbiased surveys, 
suggesting a high success rate of redshift determinations at low-redshifts,
although the statistical error is still large. 
Furthermore, the obtained metallicities do not show significant systematic trend 
with respect to burst energy ($E_{\rm iso}$) and redshift. 
Thus there is no sign that the low-redshift sample 
of the GRB host galaxies is biased compared to 
the actual population of GRB host galaxies, 
although the tests are not robust yet. 

On the other hand, the failure of the redshift determination 
of GRB 111225A in the immediate follow up observations after the burst
suggests that faint (and thus low-metallicity) host galaxies 
might be systematically missed in the redshift selected sample. 
It is also possible that a few low-redshift galaxies which are aligned 
in the foreground of higher-redshift GRBs contaminate 
the low-redshift sample as in the case of GRB 020819B. 
The more precise estimation of 
the GRB rate density and their properties, 
including those of high-redshift bursts 
which might contaminate low-redshift samples, 
will enable us to evaluate sampling biases in the redshift selected sample robustly. 

\bigskip
%% acknowledgement
We thank Gemini and Subaru observatory staffs 
for their kind support for the observations. 
YN is supported by the Research Fellowship for Young Scientists 
from the Japan Society for the Promotion of Science (JSPS).
This research has made use of the GHostS database (www.grbhosts.org), 
which is partly funded by Spitzer/NASA grant RSA Agreement No. 1287913.
Thanks are also due to Nathaniel R. Butler, Jochen Greiner, 
and Daniel A. Perley, for producing the useful online databases. 
We would also like to thank the anonymous referee for helpful comments. 

\appendix
\section*{Metallicity determination of each host galaxy}
\label{sec:metaleach}

\begin{itemize}
\item {\bf GRB 980425 host galaxy}: 
the N2 method indicates ${\rm 12+log_{10}(O/H) = 8.57\pm0.10}$ and log$_{10}\ q$ = 7.3, 
while the $R_{23}$ method does not have any valid solution. 
The [O\emissiontype{III}]$\lambda$4959 flux is not measured in \citet{Christensen:2008a}, 
and we assume theoretically expected line ratio of 
[O\emissiontype{III}]$\lambda$4959/[O\emissiontype{III}]$\lambda$5007 = 1/3 
(e.g., \cite{Rosa:1985a, Storey:2000a}). 
The lack of $R_{23}$ solution generally suggests ${\rm 12+log_{10}(O/H)} \sim$ 8.1--8.7, 
which is consistent with the N2 solution in this case. 
We adopt the N2 solution as the best estimate of the metallicity of this host galaxy. 

\item {\bf GRB 060505 host galaxy}: 
the N2 method indicates ${\rm 12+log_{10}(O/H) = 8.72\pm0.10}$ 
and log$_{10}\ q$ = 7.2 (the best estimate). 
The $R_{23}$ method does not have valid solution.

\item {\bf GRB 080517 host galaxy}: 
the emission line strength is presented as equivalent widths in \citet{Stanway:2015a}. 
We derive the emission line fluxes assuming the continuum flux density to be 
0.8, 1.7, and 1.5 $\times 10^{-16}$ erg s$^{-1}$cm$^{-2}$\AA$^{-1}$ at the wavelength 
of [O\emissiontype{II}], H$\beta$--[O\emissiontype{III}], 
and H$\alpha$--[N\emissiontype{II}], respectively. 
The N2, $R_{23}$ upper-branch, and $R_{23}$ lower-branch solutions 
are ${\rm 12+log_{10}(O/H) = 8.83\pm0.10}$, 8.68$\pm0.10$, and 8.38$\pm0.10$, 
with log$_{10}\ q$ = 7.2, 7.2, and 7.1, respectively. 
We adopt the $R_{23}$ upper-branch solution as the best estimate.

\item {\bf GRB 031203 host galaxy}: 
the N2 method indicates ${\rm 12+log_{10}(O/H) = 8.75\pm0.10}$ (log$_{10}\ q$ = 8.6), 
and the $R_{23}$ method provides two solutions of ${\rm 12+log_{10}(O/H) = 8.48\pm0.10}$ and 8.32$\pm0.10$ 
(so called upper-/lower-branches, with log$_{10}\ q$ = 8.3 and 8.2, respectively). 

KK04 calibrated the metallicity indicators 
in a range of ionization parameter $7.0 \leq {\rm log}_{10}\ q \leq 8.2$, 
and we need to extrapolate the relation when log$_{10}\ q$ is outside of this range. 
Because the N2 index is sensitive to the ionization parameter, 
we do not consider the N2 solution for this host galaxy reliable, 
although the indicated super-solar metallicity suggests 
the upper-branch solution of the $R_{23}$ method 
is more likely than the lower-branch solution. 
Thus we adopt the $R_{23}$ upper-branch solution as the best estimate. 
The $R_{23}$ upper-branch is less sensitive 
to the ionization parameter than the lower-branch and the N2. 

Although the position of this galaxy on the BPT diagram 
is consistent with being a star forming galaxy (figure~\ref{fig:BPT}), 
\citet{Levesque:2010c} showed that the host galaxy might have an AGN. 
The high $q$ parameter value also supports the existence of an AGN in the galaxy. 

\item {\bf GRB 060614 host galaxy}: 
due to the non-detection of [N\emissiontype{II}] line, 
we can put only upper-limit on the N2 index (N2 $< -1.16$). 
The [O\emissiontype{III}]$\lambda$4959 line coincides with an strong night-sky emission line, 
and the flux can not be measured from the observed spectra. 
Hence we assume the theoretically predicted line ratio of 
[O\emissiontype{III}]$\lambda$4959/[O\emissiontype{III}]$\lambda$5007 = 1/3.
The upper- and lower-branch solutions of the $R_{23}$ method 
are ${\rm 12+log_{10}(O/H) = 8.66\pm0.10}$ and 8.35$\pm0.10$ 
with log$_{10}\ q$ = 7.4 and 7.3, respectively. 
Assuming log$_{10}\ q$ = 7.3--7.4, the N2 upper-limit indicates ${\rm 12+log_{10}(O/H) < 8.4}$. 
Thus we adopt the $R_{23}$ lower-branch solution as the best estimate. 

\item {\bf GRB 030329 host galaxy}: 
the N2, $R_{23}$ upper-branch, and $R_{23}$ lower-branch solutions 
are ${\rm 12+log_{10}(O/H) = 8.07\pm0.10}$, 8.75$\pm0.10$, and 8.14$\pm0.10$, 
with log$_{10}\ q$ = 7.8, 8.2, and 7.8, respectively. 
 We adopt the $R_{23}$ lower-branch solution. 

\item {\bf GRB 120422A host galaxy}: 
the N2 method indicates ${\rm 12+log_{10}(O/H) = 8.65\pm0.10}$ 
and log$_{10}\ q$ = 7.3 (the best estimate). 
The $R_{23}$ method does not have valid solution. 

\item {\bf GRB 050826 host galaxy}: 
the N2, $R_{23}$ upper-branch, and $R_{23}$ lower-branch solutions 
are ${\rm 12+log_{10}(O/H) = 8.76\pm0.10}$, 8.86$\pm0.10$, and 8.14$\pm0.10$, 
with log$_{10}\ q$ = 7.5, 7.5, and 7.3, respectively. 
We adopt the $R_{23}$ upper-branch solution. 

\item {\bf GRB 130427A host galaxy}: 
The N2, $R_{23}$ upper-branch, and $R_{23}$ lower-branch solutions 
are ${\rm 12+log_{10}(O/H) = 8.71\pm0.10}$, 8.67$\pm0.10$, and 8.33$\pm0.10$,  
with log$_{10}\ q$ = 7.4, 7.4, and 7.3, respectively. 
We adopt the $R_{23}$ upper-branch solution as the best estimate. 
However, we note that the FOCAS TOO spectrum 
from which we measure the emission line fluxes of [O\emissiontype{II}]$\lambda$3727, 
H$\beta$, and [O\emissiontype{III}]$\lambda\lambda$4959,5007, is obtained without ADC. 
Hence it is possible that the line fluxes suffer from differential atmospheric refraction. 

To evaluate the effect of this systematic uncertainty, 
we also examine the independently measured 
emission line fluxes reported in K15 (shown in table~\ref{tab:lines}). 
The K15 line fluxes are systematically weaker 
than those in the GMOS+FOCAS spectra 
possibly due to larger loss of light at the slit, 
however we can compare relative flux ratios between the lines. 

With the K15 line fluxes, the N2, $R_{23}$ upper-branch, and $R_{23}$ lower-branch solutions 
are ${\rm 12+log_{10}(O/H) = 8.51\pm0.11}$, 8.70$^{+0.11}_{0.14}$, and 8.34$^{+0.13}_{0.12}$.  
We note that the detection of the [N\emissiontype{II}] line 
is marginal in K15 (S/N = 2.5). 
Although it is difficult to determine which 
of the $R_{23}$ solutions is true only with the K15 spectrum, 
if we adopt the upper-branch solution as suggested by our GMOS spectrum, 
it agrees well with the estimate from the GMOS+FOCAS spectra. 

\item {\bf GRB 090417B host galaxy}: 
the [O\emissiontype{II}], H$\beta$, and [O\emissiontype{III}] line fluxes are not known. 
Assuming $7.2 \leq$ log$_{10}\ q \leq 7.8$ 
(the $q$ parameter range covered by the other GRB host galaxies 
in the sample excluding that of GRB 031203), 
the N2 method indicates ${\rm 12+log_{10}(O/H)}$ = 8.6--8.9. 

\item {\bf GRB 061021 host galaxy}: 
the [N\emissiontype{II}] line is not detected providing N2 $< -0.86$. 
The $R_{23}$ upper-branch and lower-branch solutions 
are ${\rm 12+log_{10}(O/H) = 8.43^{+0.14}_{-0.15}}$ and $8.51^{+0.15}_{-0.19}$, 
respectively (log$_{10}\ q$ = 7.5 in either case). 
The N2 upper-limit is not deep enough to reject either of the $R_{23}$ solutions. 
Combining the $R_{23}$ upper-branch and lower-branch solutions, 
we estimate the metallicity of this host galaxy to be 8.28--8.66. 

\item {\bf GRB 011121 host galaxy}: 
the N2, $R_{23}$ upper-branch, and $R_{23}$ lower-branch solutions 
are ${\rm 12+log_{10}(O/H) = 8.00\pm0.10}$, 8.69$\pm0.10$, and 8.33$\pm0.10$, 
with log$_{10}\ q$ = 7.2, 7.4, and 7.2, respectively. 
We adopt the $R_{23}$ lower-branch solution. 

\item {\bf GRB 120714B host galaxy}: 
the N2 method indicates ${\rm 12+log_{10}(O/H) = 8.54\pm0.11}$, 
although the detection of the [N\emissiontype{II}] line is marginal. 
The $R_{23}$ method, which is not affected by the low-S/N of the [N\emissiontype{II}] line, 
provides the upper-branch and the lower-branch solutions: 
${\rm 12+log_{10}(O/H) = 8.50\pm0.10}$ and 8.43$\pm0.10$, 
respectively (log$_{10}\ q$ = 7.5 in either case). 
We adopt the $R_{23}$ upper-branch solution, 
which agrees with the N2 solution. 

\end{itemize}

\end{document}